\documentclass[aps,prl,superscriptaddress,twocolumn,a4paper,longbibliography]{revtex4-2}

\usepackage[utf8]{inputenc}
\usepackage[T1]{fontenc}
\usepackage{amsmath,amssymb,amsfonts,amsthm}
\usepackage{mathtools}
\usepackage{graphicx}% Include figure files
\usepackage{dcolumn}% Align table columns on decimal point
\usepackage{bm}% bold math
\usepackage[table]{xcolor}
\usepackage{tikz} %for ORCID
\usepackage{dsfont}
\usepackage{float}
\usepackage{xfrac}
\usepackage{appendix}

\usepackage[normalem]{ulem}
\usepackage{siunitx}

\graphicspath{ {./figs/} }
\definecolor{myurlcolor}{rgb}{0,0,0.7}
\definecolor{myrefcolor}{rgb}{0.8,0,0}
\usepackage{hyperref}
\hypersetup{
    unicode=true, bookmarksnumbered=false, bookmarksopen=false, breaklinks=false, pdfborder={0 0 0},
    colorlinks=true,
    linkcolor=myrefcolor,citecolor=myurlcolor,urlcolor=myurlcolor
}

\theoremstyle{plain}% default

\makeatletter
\DeclareMathSymbol{\shortminus}{\mathbin}{AMSa}{"39}

\newcommand{\sectionPRL}[1]{\noindent\textit{#1.---}\nolinebreak}

\newcommand{\mrm}[1]{\mathrm{#1}}

\newcommand{\trm}[1]{\textrm{#1}}

\newcommand{\dd}{\mathrm{d}}
\newcommand{\ee}{\mathrm{e}}

\newcommand{\TT}{\mrm{T}}

\newcommand{\Tr}{\mathrm{Tr}}
\usepackage{xparse}
\NewDocumentCommand\trace{g}{
  \IfNoValueTF{#1}
    {\Tr}
    {\Tr\!\left\{#1\right\}}
}
\newcommand{\E}[2]{\mathbb{E}_{#2}\!\left[#1\right]}
\newcommand{\EE}[1]{\mathbb{E}\!\left[{#1}\right]}

\newcommand{\eref}[1]{(\ref{#1})}
\newcommand{\eqnref}[1]{Eq.~(\ref{#1})}
\newcommand{\eqnsref}[2]{Eqs.~(\ref{#1}-\ref{#2})}
\newcommand{\figref}[1]{Fig.~\ref{#1}}

\newcommand{\appref}[1]{App.~\ref{#1}}

\newcommand{\citeref}[1]{Ref.~\cite{#1}}
\newcommand{\refcite}[1]{Ref.~\cite{#1}}

\newcommand{\dt}{\dd t}

\newcommand{\bigcdot}{ \scalebox{1.5}{\raisebox{-0.25ex}{$\,\boldsymbol{\cdot}\,$}}}

\newcommand{\Lin}{\mathcal{L}}

\newcommand{\LinOp}{\hat{L}}
\newcommand{\UnitOp}{\hat{U}}

\newcommand{\Jx}{\hat{J}_x}
\newcommand{\Jy}{\hat{J}_y}
\newcommand{\Jz}{\hat{J}_z}

\newcommand{\X}{\hat{X}}
\newcommand{\Y}{\hat{Y}}
\newcommand{\Z}{\,\hat{Z\;}\!}

\newcommand{\Fisher}{F\,}
\newcommand{\BigP}{\mathbb{P}}
\newcommand{\mapS}{\mathcal{S}}

\newcommand{\estX}{\est{\hat{X}}}
\newcommand{\estY}{\est{\hat{Y}}}

\newcommand{\sz}{\hat{\sigma}_z}
\newcommand{\cc}{\scriptscriptstyle \mathrm{(c)}} %conditional evolution for 
\newcommand{\rhoc}{\rho_{\cc}}

\newcommand{\est}[1]{\tilde{#1}}
\newcommand{\brkt}[1]{\langle #1 \rangle}
\newcommand{\brktc}[1]{\big\langle #1 \big\rangle_{\!\cc}}
\newcommand{\kcoll}{\kappa_\text{coll}}
\newcommand{\kloc}{\kappa_\text{loc}}
\newcommand{\kq}{\kappa}

\newcommand{\Vcoll}{\text{V}_{\!\text{coll}}}
\newcommand{\Vp}{\text{V}_{\!\text{P}}}
\newcommand{\Vq}{\text{V}_{\!\text{Q}}}
\newcommand{\Vloc}{\text{V}_{\!\text{loc}}}

\newcommand{\zetai}{\zeta^{\,(i)}}
\newcommand{\avgzeta}{\overline{\zeta}}
\newcommand{\fzeta}{f(\pmb{\zeta})}
\newcommand{\fzetaindex}[1]{f(\pmb{\zeta}_{#1})}

\newcommand{\I}{\mathds{1}}
\newcommand{\D}{\mathcal{D}}
\renewcommand{\H}{\mathcal{H}}
\newcommand{\Vx}{\mrm{V}_{\!X}^{\cc}}
\newcommand{\Vy}{\mrm{V}_{\!Y}^{\cc}}
\newcommand{\Vz}{\mrm{V}_{\!Z}^{\cc}}
\newcommand{\estVx}{\est{\mrm{V}}_X}
\newcommand{\estVy}{\est{\mrm{V}}_Y}
\newcommand{\estVz}{\est{\mrm{V}}_Z}
\newcommand{\estCxy}{\est{\mathrm{C}}_{XY}}

\newcommand{\Cxy}{\mathrm{C}_{\!XY}^{\cc}}

\newcommand\scalemath[2]{\scalebox{#1}{\mbox{\ensuremath{\displaystyle #2}}}}

\newcommand{\GenOp}{\mathcal{O}}

\newcommand{\Unitary}{\mathcal{U}}

\newcommand{\Diag}[1]{\mathrm{Diag}[#1]}
\newcommand{\bigzeta}{\mathcal{Z}}

\newcommand{\Dt}{{\delta t}}
\newcommand{\dW}{\mrm{d}W}
\newcommand{\measE}{\,\hat{\!E}}

\newcommand{\FF}{\mathfrak{F}}

\newcommand{\braketop}[3]{\langle #1 | #2 | #3 \rangle}

\newcommand{\banihil}{\hat{b}}
\newcommand{\bcreat}{\hat{b}^\dagger}

\renewcommand{\vec}[1]{\boldsymbol{#1}} %vector of variables
\newcounter{appendixSectionCounter}
\renewcommand{\theappendixSectionCounter}{\Alph{appendixSectionCounter}}

\newcounter{appendixSubsectionCounter}
\renewcommand{\theappendixSubsectionCounter}{\theappendixSectionCounter.\arabic{appendixSubsectionCounter}.}

\newcounter{appendixSubsubsectionCounter}
\renewcommand{\theappendixSubsubsectionCounter}{\theappendixSubsectionCounter.\arabic{appendixSubsubsectionCounter}.}

\newcommand{\appendixsection}[1]{
    \refstepcounter{appendixSectionCounter}
    \setcounter{appendixSubsectionCounter}{0} % Reset subsection counter
    \setcounter{appendixSubsubsectionCounter}{0} % Reset subsubsection counter
    \section*{Appendix \theappendixSectionCounter: #1}
    \addcontentsline{toc}{section}{Appendix \theappendixSectionCounter: #1}
}

\newcommand{\appendixsubsection}[1]{
    \refstepcounter{appendixSubsectionCounter}
    \setcounter{appendixSubsubsectionCounter}{0} % Reset subsubsection counter
    \subsection*{\theappendixSubsectionCounter\ #1}
    \addcontentsline{toc}{subsection}{\theappendixSubsectionCounter\ #1}
}

\newcommand{\appendixsubsubsection}[1]{
    \refstepcounter{appendixSubsubsectionCounter}
    \subsubsection*{\theappendixSubsubsectionCounter\ #1}
    \addcontentsline{toc}{subsubsection}{\theappendixSubsubsectionCounter\ #1}
}

\definecolor{lime}{HTML}{A6CE39}
\DeclareRobustCommand{\orcidicon}{
	\begin{tikzpicture}
	\draw[lime, fill=lime] (0,0) 
	circle [radius=0.16] 
	node[white] {{\fontfamily{qag}\selectfont \tiny ID}};
	\draw[white, fill=white] (-0.0625,0.095) 
	circle [radius=0.007];
	\end{tikzpicture}
	\hspace{-2mm}
}
\foreach \x in {A, ..., Z}{\expandafter\xdef\csname orcid\x\endcsname{\noexpand\href{https://orcid.org/\csname orcidauthor\x\endcsname}
			{\noexpand\orcidicon}}
}
\makeatother
\begin{document}

\title{Tracking time-varying signals with quantum-enhanced atomic magnetometers}

\newcommand{\CENT}{Centre of New Technologies, University of Warsaw, Banacha 2c, 02-097 Warsaw, Poland}
\newcommand{\IFPAN}{Institute of Physics, Polish Academy of Sciences, Aleja Lotnik\'{o}w 32/46, 02-668 Warsaw, Poland}
\newcommand{\ICFO}{ICFO - Institut de Ci\`encies Fot\`oniques, The Barcelona Institute of Science and Technology, 08860 Castelldefels (Barcelona), Spain}
\newcommand{\ICREA}{ICREA - Instituci\'{o} Catalana de Recerca i Estudis Avan{\c{c}}ats, 08010 Barcelona, Spain}

\author{J\'{u}lia Amor\'{o}s-Binefa\orcidA{}}
\email{j.amoros-binefa@cent.uw.edu.pl}
\affiliation{\CENT}

\author{Morgan W. Mitchell\orcidC{}}
\affiliation{\ICFO}
\affiliation{\ICREA}

\author{Jan Ko\l{}ody\'{n}ski\orcidB{}}
\email{jankolo@ifpan.edu.pl}
\affiliation{\CENT}
\affiliation{\IFPAN}

\begin{abstract}
Quantum entanglement, in the form of spin squeezing, is known to improve the sensitivity of atomic instruments to static or slowly-varying quantities. Sensing transient events presents a distinct challenge, requires different analysis methods, and has not been shown to benefit from entanglement in practically-important scenarios such as spin-precession magnetometry (SPM). Here we adapt estimation control techniques introduced in~\href{https://doi.org/10.1103/k7nk-lrwd}{[PRX Quantum 6, 030331 (2025)]} to the experimental setting of SPM and analogous techniques. We demonstrate that real-time tracking of fluctuating fields benefits from measurement-induced spin squeezing and that quantum limits dictated by decoherence are within reach of today's experiments. We illustrate this quantum advantage by single-shot tracking, within the coherence time of a spin-precession magnetometer, of a magnetocardiography signal overlain with broadband noise.
\end{abstract}

\maketitle

%%%%%%%%%%%%%%%%%%%%%%%%%%%%%%%%%%%%%%%%%%%%%%%%%%%%%%%%%%%%%%%%%%%%%%%%%%%%%%%%%%%%%%%%%%%%
\sectionPRL{Introduction}
Since the work of~\citet{Kitagaba_Ueda_SSS} and~\citet{Wineland1994} it has been known that entangled states of spin ensembles can have lower intrinsic noise than do non-entangled states. These ``spin-squeezed states'' (SSSs) have been employed in proof-of-principle demonstrations of magnetometer sensitivity~\cite{Sewell2012, MartinCiurana2017} and atomic clock short-term stability~\cite{LerouxPRL2010b, PedrozoPenafielN2020} beyond the standard quantum limit (SQL), i.e., beyond the best possible performance with a non-entangled state in equal conditions. Such entanglement-powered performance boosts are known as ``quantum enhancement'' because they benefit from correlations that are not possible in classical models of the same systems. It was noted by~\citet{Kuzmich1998} that SSSs can arise naturally in the process of non-destructive measurement~\cite{GrangierN1998, SewellNP2013}. Both magnetometers~\cite{Sewell2012, vasilakis_generation_2015} and optical atomic clocks~\cite{LerouxPRL2010b, LodewyckPRA2009} employ non-destructive measurement, suggesting that quantum enhancement using measurement-induced squeezing may be practical in state-of-the-art instruments~\cite{BenedictoPRL2022}. Indeed, quantum enhancements in high-performance instruments~\cite{Troullinou2021, RobinsonNPhys2024} have been demonstrated. 

\begin{figure}[t!]
    \includegraphics[width=\columnwidth]{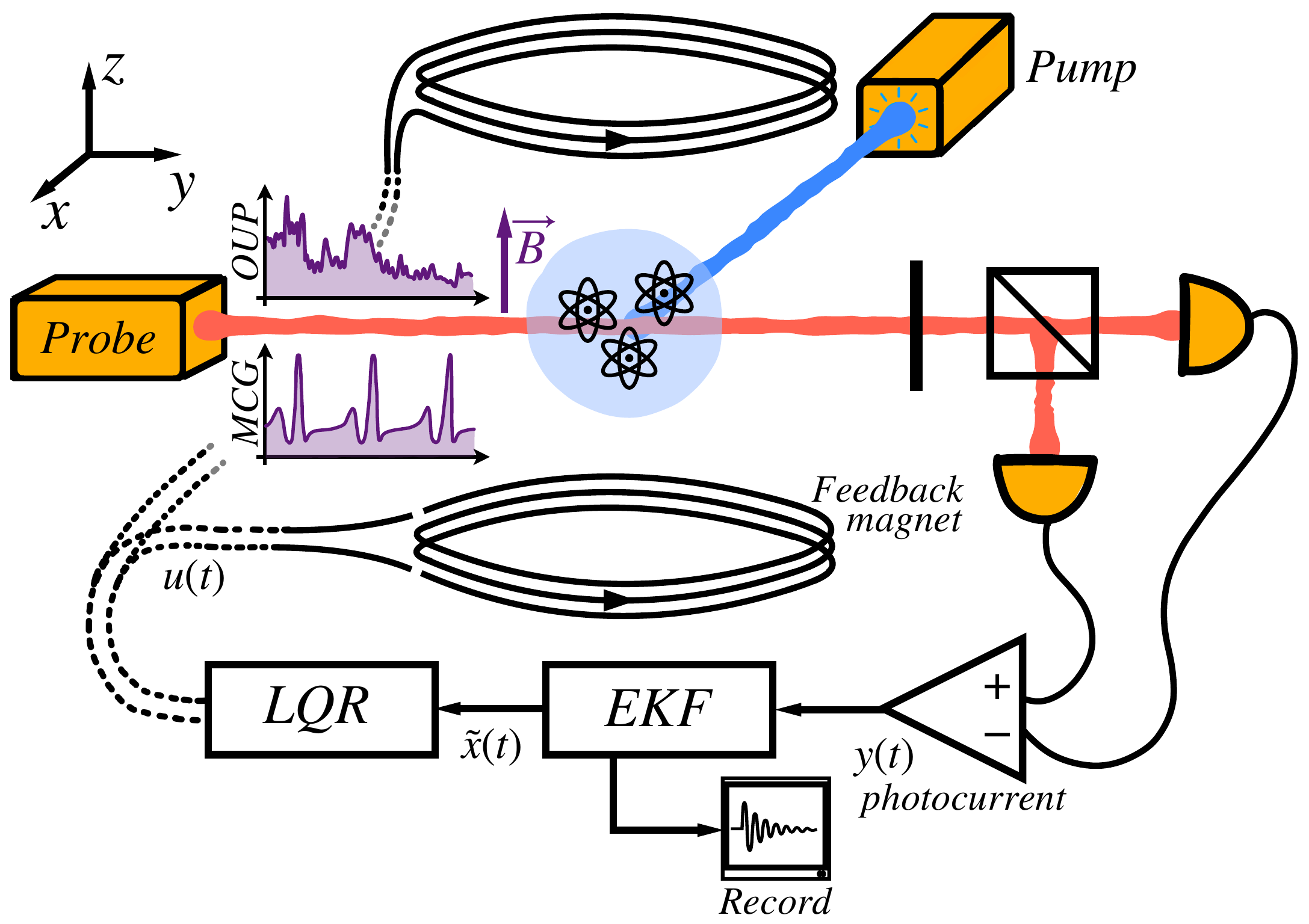}
    \caption{\textbf{Real-time atomic magnetometry}. The magnetometer consists of $N$ spin-1/2 atoms initially pumped (blue) into a coherent spin state along the $x$-axis. It is used to sense time-varying magnetic fields $B(t)$ along $z$, e.g., stochastic (OUP) or a cardiac-like (MCG) signals. This is possible by continuously probing (red) the $y$-component of the ensemble spin over time. In particular, an Extended Kalman Filter (EKF) is used to estimate in real time the field and dominant moments of the atomic spin from the detected photocurrent $y(t)$. The EKF's estimates $\tilde{x}(t)$ are then processed by the Linear Quadratic Regulator (LQR), which drives a feedback magnetic coil to produce an auxiliary field $u(t)$ that cancels any Larmor precession of the atomic spin. As a result, the sensitivity is quantum-enhanced in real time~--~the ensemble is driven into a spin-squeezed state pointing along $x$, whose variance is reduced in the $y$-direction by the measurement~\cite{Amoros-Binefa2024}.
    }
    \label{fig:setup}
\end{figure}

To date, much work on this topic has involved step-wise protocols, in which state preparation, evolution, detection, and estimation are distinct steps~\cite{GiovannettiPRL2006}. Meanwhile, many real-world sensors operate continuously -- such that these processes occur simultaneously -- and are used in real-time control or monitoring applications. Practical quantum enhancement in such scenarios requires both open-system descriptions of continuous-time quantum dynamics, and analysis and control strategies that are practical to implement in real time. Describing the system dynamics in real time formally requires a quantum model of measurement backaction~\cite{Caves1980,Braginsky1996}, i.e., of the evolution of the quantum system conditioned on the measurement record~\cite{Belavkin1989, Wiseman1993, handel_modelling_2005}. This has been achieved in experiments with Gaussian systems such as mechanical oscillators~\cite{wieczorek_optimal_2015, rossi_observing_2019, Wilson2015, Rossi2018, Magrini2021}, which have the advantage that both the conditional and unconditional distributions remain Gaussian. For spin-precession sensors, theoretical models have been proposed that assume Gaussianity~\cite{Geremia2003, Madsen2004, Molmer2004, Albarelli2017}, or resort to brute-force numerics for low atomic numbers~\cite{rossi_noisy_2020}. In experiments, however, spin-precession sensors naturally evolve toward non-Gaussian unconditional distributions, and to date Gaussian
models have been successfully applied only when ignoring~\cite{Jimenez2018} or evading~\cite{Kong2020, Troullinou2021, Troullinou2023} the measurement backaction.

Here, we show how spin-squeezing and quantum enhancement can arise in continuously-measured precessing spin ensembles in conditions that produce highly non-Gaussian unconditional distributions. We adapt the conditional dynamical model of \refcite{Amoros-Binefa2024}, in particular its ``co-moving Gaussian picture,'' to design an effective estimation and control scheme, illustrated in \figref{fig:setup}. To have realistic numbers, we apply this model to magnetometry in the conditions of \refcite{Kong2020}, which demonstrated measurement-induced generation of entanglement in an unpolarised ensemble of warm rubidium atoms.

We show that such a system, when appropriately polarized by optical pumping, can be used as a quantum-enhanced magnetometer to track time-varying fields. In particular, we focus on tracking signals that are:~(i)~fluctuating stochastically, or (ii)~determined by a continuously varying waveform, a magneto-cardiogram (MCG), which is distorted by stochastic noise that should be filtered out rather than tracked. In order to verify the optimality of our estimation and control scheme in the former case, we derive the quantum limit imposed by atomic dephasing and field fluctuations, which applies to any scheme that may involve measurement-based feedback~%
\footnote{
    This unifies previous quantum limits derived for fluctuating fields under collective dephasing~\cite{Amoros-Binefa2021} and for real-time estimation of a constant field~\cite{Amoros-Binefa2024}. 
}.
In particular, it also applies when tracking a stochastic field following an Ornstein-Uhlenbeck process (OUP)~\cite{Stockton2004,Petersen2006}.

%%%%%%%%%%%%%%%%%%%%%%%%%%%%%%%%%%%%%%%%%%%%%%%%%%%%%%%%%%%%%%%%%%%%%%%%%%%%%%%%%%%%%%%%%%%%
\sectionPRL{Atomic sensor model}
We consider the dynamical model of~\refcite{Amoros-Binefa2024}, which incorporates quantum backaction and decoherence and, hence, is capable of describing experiments such as~\refcite{Kong2020}. It applies to the canonical setup depicted in \figref{fig:setup}, in which an ensemble of atoms, whose number $N$ is constant but may fluctuate from shot to shot,
is optically pumped along the $x$-direction (blue beam in \figref{fig:setup}). For simplicity, we treat this as a collection of $N$ spin-$1/2$ particles initially prepared in a coherent spin state (CSS)~\cite{Ma2011}, whose dynamics is described by the collective angular momentum operators $\hat{J}_\alpha = \sum_{i = 1}^N \hat{\sigma}_\alpha^{(i)}/2$ with $\alpha = x, y, z$. The magnetic field we aim to track, aligned along $z$, induces precession of the ensemble spin around $z$ at an angular (Larmor) frequency $\omega(t)=\gamma B(t)$, where $\gamma$ is the gyromagnetic ratio. The atoms are continuously monitored by the probe along the $y$-axis (red beam in \figref{fig:setup}) via the Faraday effect~\cite{deutsch_quantum_2010}, which yields a continuous non-demolition measurement of $\hat{J}_y$ in a homodyne-like form~\cite{Wiseman1993}, i.e.~with the detected photocurrent:
\begin{equation} \label{eq:photocurrent}
    y(t) \dt = 2\eta \sqrt{M N} \brktc{\Y(t)}\dt + \sqrt{\eta} \, \dW,
\end{equation}
where we define $\Y \coloneqq \hat{J}_y/\sqrt{N}$ (and analogously for $\X,\Z$) with the mean $\brktc{\Y(t)} = \Tr{\{\rhoc(t) \Y\}}$ taken w.r.t.~the conditional atomic state $\rhoc(t)$;~$\eta$~is the detection efficiency, $M$ the measurement strength, and $\dW$ the Wiener increment satisfying $\EE{\dW^2}=\dt$~\cite{Gardiner1985}.

After the atoms are initialised by optical pumping into a CSS polarised along $x$ (s.t.~$\brktc{\Jx(0)}=J_0 = N/2$), they evolve according to the stochastic master equation (SME)~\cite{Belavkin1989,handel_modelling_2005}
that %models the dynamics of the conditional atomic state $\rhoc (t)$ 
takes the form~\cite{Amoros-Binefa2024}:
\begin{align}
    \dd\rhoc = &-i \left\{\omega(t) + u(t)\right\} \sqrt{N} [\!\Z\!,\rhoc] \dt \nonumber \\
    &+ \frac{\kloc N}{2} \!\! \sum_{j=1}^N \! \D[\hat{z}^{(j)}] \rhoc \dt \, + \kcoll N \D[\!\Z\!]\rhoc \dt  \nonumber \\
    &+ M \! N\D[\Y]\rhoc \dt + \sqrt{\eta M \! N} \H[\Y] \rhoc \dW, 
    \label{eq:SME}
\end{align} 
where the superoperators $\D$ and $\H$ are defined for any operator $\GenOp$ and state $\rho$ as \mbox{$\D[\GenOp]\rho \coloneqq  \GenOp \rho  \GenOp^\dagger - \frac{1}{2}\{\GenOp^\dagger\GenOp,\rho\}$} and \mbox{$\H[ \GenOp] \rho \coloneqq  \GenOp \rho + \rho  \GenOp^\dagger - \trace{(\GenOp+\GenOp^\dagger)\rho}\rho$}. 
The free evolution above describes the precession of atoms due to the magnetic field being tracked, $\omega(t) \propto B(t)$, and the instantaneous control $u(t) \equiv u(\pmb{y}_{\le t})$ that may depend on the full photocurrent record $\pmb{y}_{\le t} \coloneqq \{y(\tau)\}_{0\leq \tau \leq t}$, both applied along $z$. The model includes local (with rate $\kloc$) and collective (with rate $\kcoll$) decoherence terms that correspond to dephasing aligned with the magnetic field. The last two terms in \eqnref{eq:SME} account for the measurement backaction along $y$:~dephasing induced by the detection process and a stochastic jump correlated with the photocurrent~\eqnref{eq:photocurrent}, $\dW = [y(t)/\sqrt{\eta} - 2 \sqrt{\eta M N} \brktc{\Y}]\dt$. Parametrising the decay of the ensemble polarisation in the $xy$-plane by the spin-decoherence time $T_2$~\cite{Wang2005}, i.e.~as $\brkt{\hat{J}_{x}(t)}=J_0\exp(-t/T_2)$ in the Larmor-precessing frame, the dynamics \eref{eq:SME} yields $T_2 = 1/(\kcoll/2 + \kloc)$.

In order to reproduce conditions of \refcite{Kong2020} in absence of spin-exchange collisions, we allow the atomic number $N$ to fluctuate around $\bar{N}=10^{13}$ with 1\%-error ($N\sim\mathcal{N}(\bar{N},\sigma^2)$ with $\sigma=10^{11}$). We also set $T_2=\SI{10}{\milli\second}$, $\kcoll = 0$, $\kloc = 1/T_2 = \SI{100}{\hertz}$ and the nominal Larmor frequency to $\bar{\omega}=2\pi \times \SI{30}{\kilo\hertz}$. Not to obscure the atomic sources of noise, we assume here perfect detection efficiency $\eta = 1$. As detailed in \appref{sec:exp_param}, we consider the measurement strength $M = \SI{e-8}{\hertz}$, which depends, e.g., on the intensity and detuning of the probe. Consistently, this yields the backaction timescale as $1/(MN) = \SI{0.01}{\milli\second}$~\cite{Kong2020} (see also \appref{sec:exp_param}). Although \refcite{Kong2020} assumed $T_2$ to be dictated solely by single-atom decoherence, we later add an extra collective contribution $\kcoll = \SI{10}{\micro\hertz}$ to study also the regime in which the measurement backaction cannot produce spin-squeezing ($\kcoll > M$)~\cite{Amoros-Binefa2021}.

%%%%%%%%%%%%%%%%%%%%%%%%%%%%%%%%%%%%%%%%%%%%%%%%%%%%%%%%%%%%%%%%%%%%%%%%%%%%%%%%%%%%%%%%%%%%
\sectionPRL{Simulation with real-time estimation and feedback}
\label{sec:sim_and_inference} 
Although the exact simulation of the SME \eref{eq:SME} can be performed only for low atomic numbers, for large atomic ensembles (i.e.~$N\approx10^{13}$~\cite{Kong2020}) the dynamics of first and second moments of angular momentum operators can be reproduced by resorting to the so-called ``co-moving Gaussian approximation''~\cite{Amoros-Binefa2024}. In such a case, \eqnsref{eq:photocurrent}{eq:SME} can be rewritten, see~\appref{ap:model_form}, as a state-space model~\cite{crassidis2011optimal}:
\begin{subequations}
\label{eq:state-space_model}
\begin{align}
    &\dfrac{\dd \pmb{x}}{\dt} = f(\pmb{x},u,t,\pmb{\xi}) \label{eq:model_state}\\
    &y(t) = h(\pmb{x},t) + \xi  = H \pmb{x} + \xi, \label{eq:model_measurement}
\end{align}
\end{subequations}
where the state vector $\pmb{x} =\pmb{x}_{at} \oplus \pmb{x}_{s}$ comprises relevant atomic degrees of freedom, i.e.,~conditional means of angular momenta and their corresponding (co)variances that affect the dynamics~\cite{Amoros-Binefa2024}, $\pmb{x}_{at}= (\brktc{\X},\brktc{\Y},\Vx,\Vy,\Vz,\Cxy)^T$;~and the signal encoded in the field being tracked, e.g.,~its Larmor frequency $\pmb{x}_{s}\!=\!(\omega(t))^T$. Similarly, the noise vector $\pmb{\xi}=(\xi,\xi_\omega)^T$ may be split into stochastic (Langevin) terms present in the dynamics, i.e.~the measurement backaction noise $\xi = \dW/\dt$ affecting \eqnsref{eq:photocurrent}{eq:SME} in a correlated manner; and noises affecting the estimated signal itself, i.e.~$\xi_\omega = \dW_\omega/\dt$ representing white-noise fluctuations of the field. {For the} exact analytical form of the non-linear function describing the evolution of the state, $f(\pmb{x},u,t,\pmb{\xi})$, along with the form of the $H$-matrix, see~\appref{ap:model_form}.

As described in \figref{fig:setup}, our estimation and control scheme relies on constructing estimates of $\pmb{x}(t)$ in real time with help of an extended Kalman filter (EKF), denoted as $\est{\pmb{x}}(t)$, which are then used instantaneously by the linear quadratic regulator (LQR, described below) to choose the control field $u(t)$ in \eqnref{eq:SME}. The EKF estimator is found by integrating the following differential equations along the photocurrent record $\pmb{y}_{\le t}$~\cite{crassidis2011optimal}:
\begin{subequations}
\label{eq:EKF_alg}
    \begin{align}
        \dot{\est{\pmb{x}}} &= \pmb{f}(\est{\pmb{x}},u,0,t) + K (y(t)-h(\est{\pmb{x}},0,t)) 
        \label{eq:EKF_dyn}\\
        \dot{\Sigma} &= (F - G S R^{-1} H)\Sigma + \Sigma(F - G S R^{-1} H)^\TT + \nonumber\\
        & \; \; \; \, + G(Q - S R^{-1} S^\TT) G^\TT - \Sigma H^\TT R^{-1} H \Sigma,
        \label{eq:Sigma_dyn}
    \end{align}
\end{subequations}
where the update-predict equation for the estimate \eqref{eq:EKF_dyn} is coupled to the Riccati equation describing the evolution of the EKF covariance $\Sigma(t)$ in \eqnref{eq:Sigma_dyn} via the Kalman gain $K \coloneqq (\Sigma H^\TT - G S) R^{-1}$. The form of (Jacobian) dynamical matrices $F(t)$ and $G(t)$, as well as $H$, is determined by the model \eref{eq:state-space_model}:~$F(t) \coloneqq \nabla_{\pmb{x}} \pmb{f}|_{(\est{\pmb{x}},u,0)}$, $G(t) \coloneqq \nabla_\xi \pmb{f} |_{\est{\pmb{x}}}$ and $H \coloneqq \nabla_{\pmb{x}} h$;~however, as their exact entries depend on most recent estimates, $\est{\pmb{x}}(t)$, these must be reevaluated at each step of the EKF algorithm \eref{eq:EKF_alg}. In contrast, the noise matrices $Q \coloneqq \E{\pmb{\xi} \, \pmb{\xi}^\TT}{} = \Diag{1,q_K}$ with $q_K > 0$, $R \coloneqq \E{\zeta^2}{} = \eta$, and $S \coloneqq \E{\pmb{\xi} \zeta}{}= (\sqrt{\eta},\, 0)^\TT$ are predetermined, whereas the initial estimates, $\est{\pmb{x}}(0)$, and their covariance (initial error), $\Sigma(0) = \EE{\Delta^2 \est{\pmb{x}}(0)}$, are set according to the prior distribution assumed~\cite{crassidis2011optimal}. 

For non-linear models, such as \eref{eq:state-space_model}, the EKF covariance $\Sigma(t)$ is not guaranteed to match the true error, i.e.~the {average} mean squared error (aMSE) matrix $\EE{\Delta^2 \est{\pmb{x}}(t)} \coloneqq \E{(\est{\pmb{x}}(t) - \pmb{x}(t))(\est{\pmb{x}}(t) - \pmb{x}(t))^\TT}{p(\pmb{y}_{\leq t},\pmb{x}(0))}$. However, as here we simulate and, hence, have access to the true state $\pmb{x}(t)$, we may explicitly verify whether $\Sigma(t)$ provides a faithful prediction of $\EE{\Delta^2 \est{\pmb{x}}(t)}$ by averaging the latter over sufficient measurement records---in particular, for the aMSE of the tracked signal $\Delta^2 \est{\omega}(t)$.

To decide on the control strategy, we consider the field-compensating regime, $u(t)\approx-\omega(t)$, in which the atomic spin can be approximated by a bosonic mode~\cite{Madsen2004,Molmer2004}. Defining then a quadratic cost that quantifies deviations of the spin from pointing along $x$, it is minimised by choosing the control field to evolve under LQR equations similar to \eqref{eq:EKF_alg}~\cite{crassidis2011optimal}, obtained upon linearising the function $f$ in \eqnref{eq:model_state}~\cite{Stockton2004,Amoros-Binefa2024}. Within the steady-state regime, the form of the optimal control is then given by $u(t)= - \est{\omega}(t) - \lambda \, \brktc{\estY(t)}$, where $\lambda$ is a gauge parameter determined by the cost normalisation---we set $\lambda=1$ throughout this work---whereas $\est{\omega}(t)$, $\brktc{\estY(t)}$ are real-time estimates of the Larmor frequency and the spin $y$-component, respectively, provided by the EKF $\est{\pmb{x}}(t)$.

\begin{figure}[t!]
    \includegraphics[width=\columnwidth]{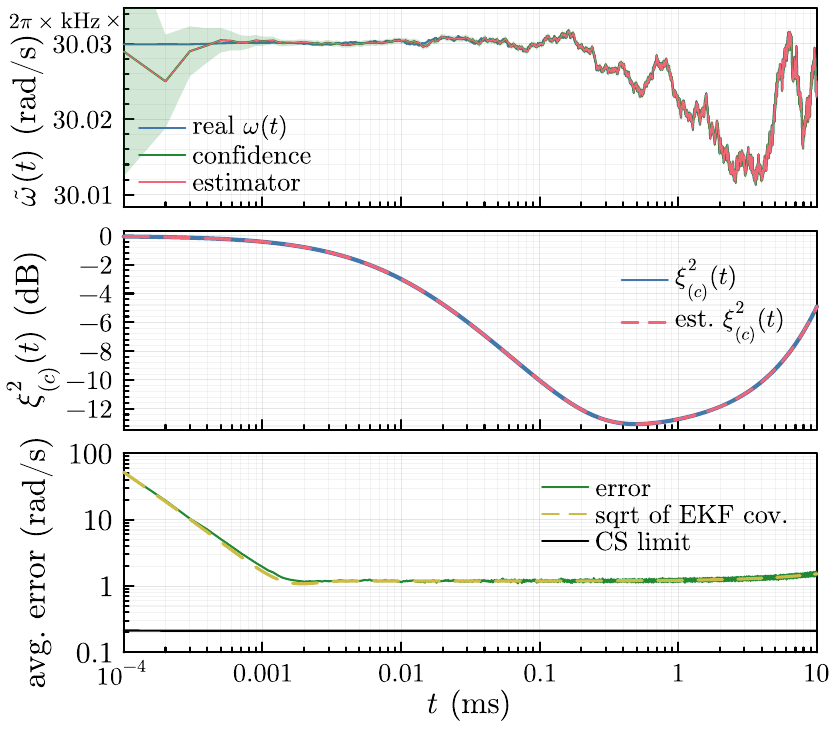}
    \caption{%
    \textbf{Quantum-enhanced tracking of a fluctuating magnetic field.} 
    \emph{Top}:~Stochastic fluctuations of the field ({blue}) being efficiently tracked in real time by the EKF estimate (red) after gathering only $\approx\!\!\SI{0.01}{\milli\second}$ data of the photocurrent. The shaded area represents the confidence band limited by $\pm 2 \sqrt{\EE{\Delta^2 \est{\omega}(t)}}$, i.e.~the error obtained upon averaging.
    \emph{Middle}:~Evolution of the (average) spin-squeezing parameter ({blue}), compared with its (average) real-time prediction by the EKF ({red}).
    \emph{Bottom}:~Evolution of the average error ({green}) in estimating the fluctuating field, $\sqrt{\EE{\Delta^2 \est{\omega}(t)}}$, which stabilises at the value $\approx\!\! \SI{1}{\radian\,\second^{-1}}$, as correctly predicted by the EKF covariance ({dashed, yellow}). The quantum limit imposed by local dephasing ({black}) is not attained due to insufficient measurement strength (here $M=\SI{e-8}{\hertz}$~\cite{Kong2020}, but see also Appendix \figref{fig:largerM} for larger $M$)~\cite{Amoros-Binefa2024}. In all plots, averaging was performed over 1000 field+atom stochastic trajectories.
    }
    \label{fig:oup_sq_estimation}
\end{figure}

%%%%%%%%%%%%%%%%%%%%%%%%%%%%%%%%%%%%%%%%%%%%%%%%%%%%%%%%%%%%%%%%%%%%%%%%%%%%%%%%%%%%%%%%%%%%
\sectionPRL{Tracking fluctuating magnetic fields} 
\label{sec:tracking_OUP}
We first demonstrate efficient tracking of field fluctuations {that evolve under} an Ornstein-Uhlenbeck process (OUP)~\cite{Stockton2004,Petersen2006}:
\begin{equation} \label{eq:oup}
    \dd \omega(t) = -\chi ( \omega(t) - \bar{\omega} )\dt + \sqrt{q_\omega} \dW_\omega,
\end{equation}
where $\chi>0$ is the decay parameter and $\bar{\omega}$ is the equilibrium value towards which $\omega(t)$ reverts. The fluctuations with a strength of $q_\omega>0$ are introduced by a Wiener increment $\dW_\omega$ of variance $\EE{\dW_\omega^2} = \dt$. We set here $\chi = \SI{1}{\second^{-1}}$ and $q_\omega = 10^6 \, \SI{}{\radian^{2}\second^{-3}}$ so that the field is perturbed on average by $\approx\!0.05\%$ from the nominal angular frequency $\bar{\omega}=2\pi \times \SI{30}{\kilo\hertz}$ over the magnetometer coherence time $T_2=\SI{10}{\milli\second}$ (recall that an OUP exhibits variance of $\approx\! q_\omega t$ at short times $t \lesssim 1/\chi$~\cite{Amoros-Binefa2021}). Importantly, we generalise our previous results~\cite{Amoros-Binefa2021, Amoros-Binefa2024} to establish the quantum limit on the aMSE, $\Delta^2 \est{\omega}(t)$, for fields fluctuating according to the OUP \eref{eq:oup}, which applies to all magnetometry setups of \figref{fig:setup} that involve \emph{any} form of continuous measurement and measurement-based feedback.  The so-called ``Classical Simulation'' limit~\cite{matsumoto_metric_2010,Demkowicz2012} is dictated by the presence of dephasing terms with rates $\kloc$ and $\kcoll$ in the SME \eref{eq:SME}, and also accounts for the stochastic character of the field (see \appref{ap:CSbound} for details):
\begin{align} \label{eq:CSlimit_main}
    \!\!\!\E{\EE{\Delta^{\!2}\est{\omega}(t) }}{p(N)}  
    \!\geq\! \sqrt{q_\omega \kq(\bar{N})} \coth{\left(t \sqrt{\frac{q_\omega}{\kq(\bar{N})}}\right)},
\end{align}
where by $\kq(\bar{N}) \coloneqq \kcoll + 2\kloc/\bar{N}$ we define the effective dephasing rate arising due to both  collective and local contributions. In fact, the r.h.s.~of \eqnref{eq:CSlimit_main} applies for any atomic number $N$ and so, by convexity arguments, see \appref{ap:CSbound}, it also applies to its mean value $\bar{N}$.

In \figref{fig:oup_sq_estimation}, using the experimental parameters of \refcite{Kong2020}, we demonstrate that the EKF successfully tracks the fluctuating field in real-time (\emph{top}). Moreover, it provides an accurate estimate of the atomic spin-squeezing parameter~\cite{Ma2011} (\emph{middle}), defined as $\xi^2(t) \coloneqq N \Vy(t)/\brktc{\Jx(t)}^2$, which reaches $\SI{-13}{\decibel}$ at around \SI{0.5}{\milli\second}%
~\footnote{
    The value of $\approx\!\SI{-2}{\decibel}$ reached in \refcite{Kong2020} should not be directly compared, as therein an unpolarised ensemble in the spin-exchange relaxation-free regime was considered. 
}. This squeezing, induced purely by the measurement backaction, consistently emerges at 
$\approx\!\SI{0.01}{\milli\second}$ timescales. At longer times ($\SI{0.01}{\milli\second}\lesssim t \lesssim T_2$), the magnetometer reaches its optimal resolution, enabling real-time tracking of field fluctuations with an error of $\approx\SI{1}{\radian \, \second^{-1}}$ in real time (\emph{bottom}). The minimal aMSE (green) is correctly predicted by the EKF covariance (orange, dashed). Although the quantum limit \eref{eq:CSlimit_main} imposed by local dephasing ($\kcoll=0$) sets a fundamental error (black) of $\approx\SI{0.045}{\radian \, \second^{-1}}$, this limit is not reached in the setup considered. However, it could, in principle, be attained by increasing the measurement strength $M$~\cite{Amoros-Binefa2024}, as shown in Appendix \figref{fig:largerM}.

\begin{figure}[t!]
    \includegraphics[width=\columnwidth]{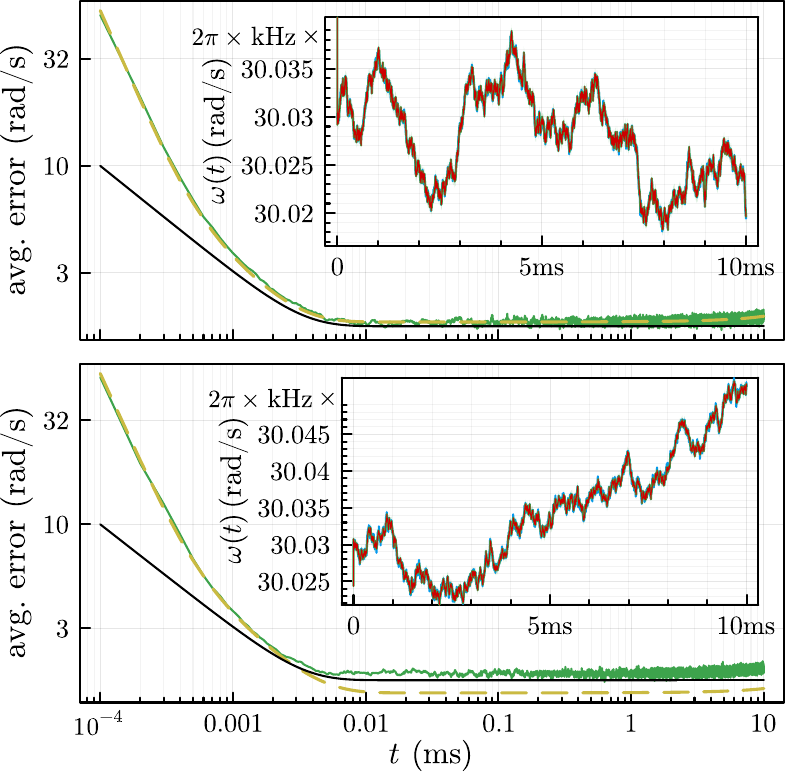}
    \caption{\textbf{Tracking field fluctuations at the quantum limit \eref{eq:CSlimit_main}}. As in the bottom plot of \figref{fig:oup_sq_estimation}, the (true) average error, $\sqrt{\EE{\Delta^2 \est{\omega}(t)}}$, is depicted against the error predicted by the EKF (dashed, yellow) and the limit \eref{eq:CSlimit_main} imposed by the decoherence (black), but this time including a tiny contribution of collective dephasing $\kcoll = \SI{10}{\micro\hertz}$~\cite{Note3}. The measurement strength achieved in~\cite{Kong2020} is now sufficient for the magnetometer to operate at the quantum limit \eref{eq:CSlimit_main}, no matter whether the EKF is provided with the exact OUP dynamics \eref{eq:oup} of the field (\emph{top}) or its mismatched version (\emph{bottom}). Although in the latter case the EKF expects fluctuations of twice smaller strength ($q_K = q_\omega/2$) and much faster decay ($\chi_K = 10\chi$), the average error ($\sqrt{\mrm{aMSE}}$, green) still attains the quantum limit (black), while the EKF covariance (dashed, yellow) underestimates the error. Both plots were obtained by averaging over 1000 field+atom stochastic trajectories. In each case, the \emph{inset} presents an exemplary field trajectory (blue) together with its EKF real-time estimate (red), which is well within the confidence interval $\pm 2\sqrt{\EE{\Delta^2 \est{\omega}(t)}}$ (shaded green).
    }
    \label{fig:oup_estimation}
\end{figure}

As the quantum limit \eref{eq:CSlimit_main} is dictated by both collective and local dephasing, in \figref{fig:oup_estimation} we now add also a tiny contribution of collective noise, i.e.~set $\kcoll = \SI{10}{\micro\hertz}$ ($1/\kcoll \!\approx\!\SI{1}{day}$) apart from $\kloc=\SI{100}{\hertz}$, which hardly affects the magnetometer minimal error~%
\footnote{
    Note, however, that spin-squeezing is now impossible as $\kcoll>M$~\cite{Amoros-Binefa2021}.
}, 
remaining at the $\approx\!\SI{1.94}{\radian \, \second^{-1}}$ error-level (green), but raises the limit \eref{eq:CSlimit_main} to $\approx\!\SI{1.78}{\radian \, \second^{-1}}$ (black). As a result, the magnetometer operates at its full capacity without the need to increase the measurement strength~\cite{Kong2020}, while the limit \eref{eq:CSlimit_main} is attained not only when the EKF is provided with the exact OUP field-dynamics \eref{eq:oup} (\emph{top}), but also when tracking mismatched field fluctuations (\emph{bottom}). In the latter case, the EKF is set according to the model \eref{eq:state-space_model} but with half as strong field fluctuations, $q_K = q_\omega/2$, whose decay is ten times faster, $\chi_K = 10\chi$. Although the EKF covariance (dashed, yellow) can no longer be trusted, the error (green) is hardly affected and still saturates the quantum limit \eref{eq:CSlimit_main} (black).

%%%%%%%%%%%%%%%%%%%%%%%%%%%%%%%%%%%%%%%%%%%%%%%%%%%%%%%%%%%%%%%%%%%%%%%%%%%%%%%%%%%%%%%%%%%%
\sectionPRL{Tracking noisy MCG-like signals}
\label{sec:tracking_MCG} 
A real-life application of atomic magnetometers is magnetocardiography (MCG)~\cite{Bison2009,jensen_magnetocardiography_2018,Kim2019,Yang2021}---a non-invasive method to image the magnetic field from the cardiac electrical activity~\cite{mcg_paper}. Its aim is to accurately recover in real time the underlying waveform of the field produced by the heart, in particular, its characteristic components:~the P-wave, the QRS-complex and the T-wave~\cite{MCG_signal}. In contrast to tracking fluctuating fields above, the stochastic character of the signal arises due to noise and should be filtered out~\cite{ECG_noise_removal}, preferably without resorting to time-averaging~\cite{mcg_paper}.

\begin{figure}[t!]
    \centering
    \includegraphics[width=\columnwidth]{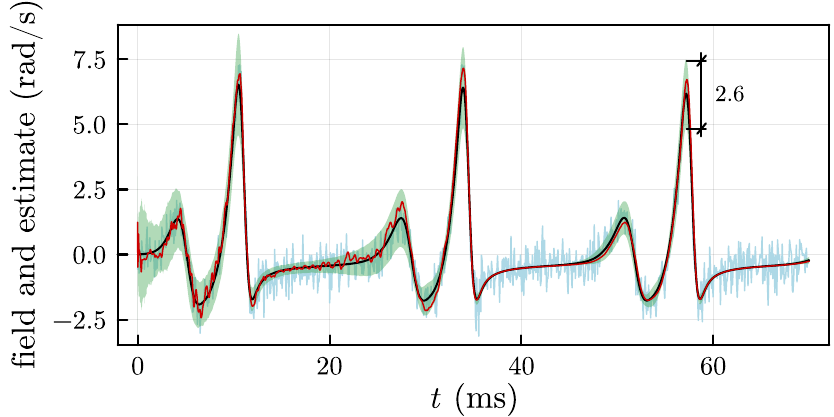}
    \caption{\textbf{Tracking a MCG-like signal.} The magnetometer, under the same conditions as in \figref{fig:oup_sq_estimation}, driven now by the field (blue) representing a noisy magnetocardiogram (MCG) in pT-range~\cite{Bison2009,Kim2019}, whose clean waveform (black) is to be tracked. The EKF within our estimation and control scheme is set to expect the signal as a solution of a VdP equation~\cite{Kaplan2008,Das2013}. The EKF estimate (red) follows the waveform very well once the magnetometer stabilises over one MCG-cycle ($\approx\!20$ms), with the highest average error ($\pm 3\sqrt{\EE{\Delta^2\est{\omega}(t)}}$ averaged over 1000 trajectories, green shading) observed at the R-wave already at $\SI{2.6}{\radian\,\second^{-1}}$ within the third cycle.
    }
    \label{fig:mcg_estimation}
\end{figure}

We demonstrate that our scheme based on the EKF+LQR feedback loop naturally accommodates waveform tracking tasks. We generate noiseless MCG-like signals (black in \figref{fig:mcg_estimation}) as dynamics of a filtered Van der Pol (VdP) oscillator~\cite{Kaplan2008,Das2013}, see details in \appref{app:VdP_details}, which we distort by adding white noise of fixed density (blue in \figref{fig:mcg_estimation}). We simulate the resulting magnetometer dynamics under experimental conditions as before~\cite{Kong2020}, upon setting the EKF parameters to match those used to generate the clean VdP signal. In \figref{fig:mcg_estimation}, we present results for a cyclic MCG-like signal of $\approx\!\SI{20}{\milli\second}$ periodicity that varies around the offset-field between minimal and maximal values of $[\SI{-2.5}{\radian/\second},\SI{7.5}{\radian/\second}]$, which correspond the magnetic range $[\SI{-56.8}{\pico\tesla},\SI{170.5}{\pico\tesla}]$ for the Rb-87 ground state hyperfine gyromagnetic ratio, compatible with human-heart fields~\cite{Bison2009,Kim2019}. Once the scheme stabilises after the first cycle, the EKF estimate (red) follows very accurately the true waveform (black), even though it is the noisy field (blue) with white-noise density $\SI{2.5e-7}{\radian^{2}\second^{-1}}$ that is the magnetometer raw signal.

%%%%%%%%%%%%%%%%%%%%%%%%%%%%%%%%%%%%%%%%%%%%%%%%%%%%%%%%%%%%%%%%%%%%%%%%%%%%%%%%%%%%%%%%%%%%
\sectionPRL{Conclusions}
\label{sec:conclusions} 
With help of a novel dynamical model capable of accurately describing quantum measurement backaction and decoherence in orientation-based (spin-1/2) atomic magnetometers~\cite{Amoros-Binefa2024}, we simulate experimental conditions (magnetic-field range, optical pumping/probing strength and spin-relaxation rate) of \citeref{Kong2020}, while disregarding effects of spin-exchange collisions, in order to show that an estimation and control scheme based on 
the EKF and LQR opens doors for tracking time-varying magnetic fields at the quantum limit. For fluctuating fields, we derive a fundamental bound on sensitivity applicable to any setup involving measurement-based feedback, and show it to be attainable by the EKF+LQR scheme. Moreover, we demonstrate that the EKF+LQR solution naturally accommodates tracking magnetocardiograms, in which case the EKF reproduces accurately the underlying waveform of a cardiac signal while filtering out the noise. Our findings provide an important step in the development of theory for atomic magnetometers to leverage spin-squeezing in real-time applications. Although we assume idealised, infinitely fast measurement---adapting our analysis to a realistic discrete-time readout is possible requiring a continuous-discrete formulation of the EKF~\cite{Dilcher2025} combined with numerical strategies allowing for precise integration of system dynamics over finite time increments~\cite{Rouchon2015}---we hope that our methods will soon be adapted to apply to orientation-based magnetometers in the Bell-Bloom configuration~\cite{Bell1961,Troullinou2021,Troullinou2023}, alignment-based magnetometers~\cite{Ledbetter2007,Weis2006,Kozbial2024};~as well as to accommodate for spin-exchange interatomic collisions~\cite{Happer1977,Savukov2005} and more general forms of atomic decoherence~\cite{Happer2010}.

\begin{acknowledgments}
We thank Marco G. Genoni, Matteo A. C. Rossi, Kasper Jensen, Klaus M{\o}lmer and Francesco Albarelli for many helpful comments. This research was funded by the National Science Centre, Poland under grant no.~2023/50/E/ST2/00457; by European Commission projects Field-SEER (ERC 101097313), OPMMEG (101099379) and QUANTIFY (101135931); Spanish Ministry of Science MCIN project SAPONARIA (PID2021-123813NB-I00) and SALVIA (PID2024-158479NB-I00),  ``NextGenerationEU/PRTR'' (Grant FJC2021-047840-I) and ``Severo Ochoa'' Center of Excellence CEX2019-000910-S;  Generalitat de Catalunya through the CERCA program,  DURSI grant No. 2021 SGR 01453 and QSENSE (GOV/51/2022).  Fundaci\'{o} Privada Cellex; Fundaci\'{o} Mir-Puig.
\end{acknowledgments}

\nocite{Albarelli2024,Wiseman1994_feedback,Wiseman1994_squeezing_feedback} % appearing in appendix

%%%%%%%%%%%%%%%%%%%%%%%%%%%%%%%%%%%%%%%%%%%%%%%%%%%%%%%%%%%%%%%%%%%%%%%%
\bibliographystyle{myapsrev4-2}
\bibliography{EKFfluct}
%%%%%%%%%%%%%%%%%%%%%%%%%%%%%%%%%%%%%%%%%%%%%%%%%%%%%%%%%%%%%%%%%%%%%%%%

%%%%%%%%%%%%%%%%%%%%%%%%%%%%%%%%%%%%%%%%%%%%%%%%%%%%%%%%%%%%%%%%%%%%%%%%%%%%%%%%%%%%%%%%%%%%%%
%%%%%%%%%%%%%%%%%%%%%%%%%%%%%%%%%%%%%%%%%%%%%%%%%%%%%%%%%%%%%%%%%%%%%%%%%%%%%%%%%%%%%%%%%%%%%%
% APPENDICES
\begin{appendices}
%%%%%%%%%%%%%%%%%%%%%%%%%%%%%%%%%%%%%%%%%%%%%%%%%%%%%%%%%%%%%%%%%%%%%%%%%%%%%%%%%%%%%%%%%%%%%%
%%%%%%%%%%%%%%%%%%%%%%%%%%%%%%%%%%%%%%%%%%%%%%%%%%%%%%%%%%%%%%%%%%%%%%%%%%%%%%%%%%%%%%%%%%%%%%

%%%%%%%%%%%%%%%%%%%%%%%%%%%%%%%%%%%%%%%%%%%%%%%%%%%%%%%%%%%%%%%%%%%%%%%%%%%%%%%%%%%%%%%%%%%%%%
\appendixsection{Model} 
\label{ap:model_form}
The non-linear function $f(\pmb{x},u,t,\pmb{\xi})$ depends on the control law $u(t) \coloneqq u(t|\pmb{y}_{\leq t})$, the noise vector $\pmb{\xi} = \pmb{\xi}_{at} \oplus \pmb{\xi}_{s}$, and the state vector $\pmb{x} =\pmb{x}_{at} \oplus \pmb{x}_{s}$. These vectors include components $\pmb{x}_{at}$ and $\pmb{\xi}_{at}$, describing atomic evolution, and components $\pmb{x}_s$ and $\pmb{\xi}_s$, accounting for the signal evolution. The components $\pmb{x}_{at}$ and $\pmb{\xi}_{at}$ are given by our atomic sensor model:~$\pmb{x}_{at} = (\brktc{\X},\brktc{\Y},\Vx,\Vy,\Vz,\Cxy)^T$ and $\pmb{\xi}_{at} = \xi$, where $\brktc{\X}$ and $\brktc{\Y}$ are the mean of the normalized collective angular momenta $\X = \Jx/\sqrt{N}$ and $\Y = \Jy/\sqrt{N}$, and $\Vx,\Vy,\Vz,$ and $\Cxy$ are the variances and covariances of such operators, i.e.~$\mrm{C}^{\cc}_{\alpha \beta}(t) \coloneqq \frac{1}{2 N} \left(\brktc{\{\hat{J}_\alpha(t),\hat{J}_\beta(t)\}}-2\brktc{\hat{J}_\alpha(t)}\brktc{\hat{J}_\beta(t)} \right)$ with diagonal elements $\mrm{V}^{\cc}_{\alpha}(t)\coloneqq \mrm{C}^{\cc}_{\alpha\alpha}(t)$ ($\alpha,\beta = X, Y, Z$). The signal components, however, depend on the specific signal being tracked and its dynamical equation. 

Similarly, the state function $f(\pmb{x},u,t,\pmb{\xi})$ can also be decomposed into two terms: one corresponding to the evolution of the atoms, and another characterizing the field, respectively. Namely, $f(\pmb{x},u,t,\pmb{\xi}) = (f_{at}(\pmb{x},u,t,\pmb{\xi}), f_s(\pmb{x},u,t,\pmb{\xi}))^T$. The state function describing the atomic evolution, $f_{at}(\pmb{x},u,t,\pmb{\xi})$, is fixed by the model of the atomic magnetometer:
\begin{align}
    f_{at}(\pmb{x}, u, t, \pmb{\xi}) = \begin{pmatrix}
        f_{\X}(\pmb{x}, u, t, \pmb{\xi}) \\
        f_{\Y}(\pmb{x}, u, t, \pmb{\xi}) \\
        f_{\Vx}(\pmb{x}, u, t, \pmb{\xi}) \\
        f_{\Vy}(\pmb{x}, u, t, \pmb{\xi}) \\
        f_{\Vz}(\pmb{x}, u, t, \pmb{\xi}) \\
        f_{\Cxy}(\pmb{x}, u, t, \pmb{\xi})
    \end{pmatrix}
\end{align}
where the form of each component of the function $f_{at}(\pmb{x}, u, t, \pmb{\xi})$ can be derived from the SME in \eqnref{eq:SME}, following the steps described in~\cite{Amoros-Binefa2024}. Namely,
\begin{align}  
    &  f_{\X}(\pmb{x}, u, t, \pmb{\xi}) \!=\! - (\omega(t)\!+\! u(t))   \brktc{\Y} \nonumber \\ 
    & \quad \quad \;\; -\! \frac{1}{2}(\kcoll \!+\! 2 \kloc \!+\! M) \brktc{\X} \nonumber \\
    &\quad \quad \;\;+\! 2\sqrt{\eta M N} \Cxy \, \xi \\ 
    & f_{\Y}(\pmb{x}, u, t, \pmb{\xi}) \!=\! (\omega(t)\!+\! u(t))  \brktc{\X}  \nonumber \\ 
    &\quad \quad \;\;- \frac{1}{2}(\kcoll \!+\! 2\kloc)  \brktc{\Y}  \nonumber \\
    &\quad \quad \;\;+\! 2\sqrt{\eta M N} \Vy \, \xi \label{eq:dJy}\\
    & f_{\Vx}(\pmb{x}, u, t, \pmb{\xi}) \! = \! - 2 (\omega(t)\! + \! u(t))  \Cxy  \nonumber \\
    & \quad \quad \;\; + \! \kcoll \! \left( \! \Vy \!+\!  \brktc{\Y}^2 \!-\! \Vx \!\right)\!   \nonumber \\
    &\quad \quad \;\; +\! \kloc \!\left( \!\frac{1}{2} \!-\!2 \Vx \!\right)\! \nonumber \\
    &\quad \quad \;\; +\! M\! \left(  \Vz \!-\! \Vx \!-\! 4 \eta N {\Cxy}^2 \, \right)\! \label{eq:dVx}\\
    &   f_{\Vy}(\pmb{x}, u, t, \pmb{\xi}) \!=\! 2 (\omega(t)\!+\! u(t))  \Cxy  \nonumber \\
    &\quad \quad \;\;+\! \kcoll \!\left(\!  \Vx \!+\!  \brktc{\X}^2 \!-\!  \Vy\!\right)\!   \nonumber \\
    &\quad \quad \;\; +\! \kloc \!\left(\! \frac{1}{2} \!-\!2 \Vy \!\!\right)\!  \!-\! 4 \eta M N {\Vy}^2  \label{eq:dVy}\\
    &   f_{\Vz}(\pmb{x}, u, t, \pmb{\xi}) \!=\!  M \!\left( \!\Vx \!+\!  \brktc{\X}^2 \!-\! \Vz\! \right)\!  \label{eq:dVz}\\
    &   f_{\Cxy}(\pmb{x}, u, t, \pmb{\xi}) \!=\! (\omega(t)\!+\! u(t)) \!\left(\!\Vx \!-\!  \Vy\!\right)\!  \nonumber \\
    &\quad \quad \;\;-\! \kcoll \!\left(2\Cxy \!+\! \brktc{\X}\brktc{\Y}\!\right)  \nonumber \\
    &\quad \quad \;\;-\! 2\kloc \Cxy \nonumber \\
    &\quad \quad \;\;-\! \frac{1}{2} M \Cxy \left( 1 + 8\eta N \Vy\right).\label{eq:dCxy} 
\end{align}
Given that the atoms are initially prepared in a coherent spin state, then the initial values for the first and second moments are $(\brktc{\X},\brktc{\Y},\Vx,\Vy,\Vz,\Cxy) = (\sqrt{N}/2,0,0,1/4,1/4,0)$.

%%%%%%%%%%%%%%%%%%%%%%%%%%%%%%%%%%%%%%%%%%%%%%%
\appendixsubsection{Ornstein-Uhlenbeck Process}
When the magnetic field  we aim to track follows an Ornstein-Uhlenbeck process just like the one in \eqnref{eq:oup}, then the component of the state vector corresponding to the signal evolution is simply $\pmb{x}_s = \omega(t)$, with a noise component $\pmb{\xi}_s = \xi_\omega = \dW_\omega / \dt$. Moreover, the part of the non-linear state space function $f(\pmb{x}, u, t, \pmb{\xi})$ modelling the dynamics of the signal becomes:
\begin{align}
    f_s(\pmb{x}, u, t, \pmb{\xi}) \equiv f_{\omega}(\pmb{x}, u, t, \pmb{\xi}) \!=\! -\chi \omega(t) + \xi_\omega
\end{align}

%%%%%%%%%%%%%%%%%%%%%%%%%%%%%%%%%%%%%%%%%%%%%%%
\appendixsubsection{Filtered Van der Pol oscillator}
\label{app:VdP_details}
For a MCG-like signal modelled by the filtered VdP oscillator, the state variables describing the signal evolution are $\pmb{x}_s = (\nu(t),\omega(t),\upsilon(t))$, as the dynamical model of the filtered VdP consists of a system of three ODEs:
\begin{align} \label{eq:VdP}
    &\dd \nu(t) = - p \, \omega(t) \,\dt \\
    &\dd \omega(t) = \frac{k}{m} \nu(t) \,\dt + 2 \frac{c}{m} (1-\upsilon(t)) \, \omega(t) \,\dt \\
    &\dd \upsilon(t) = \frac{|\nu(t)|-\nu(t)}{2 T} \,\dt - \frac{\upsilon(t)}{T} \,\dt
\end{align}
where $p, k, m, c, T > 0$ are all positive constants, and the second component, i.e. $\omega(t)$, is the variable we aim to track. The parameters specifying the MCG-like signal chosen throughout this work are: $p = 10^3$, $k = 1$, $m = 0.00098$, $c = 1$ and $T = 0.003$, with initial values: $\nu_0 = \omega_0 = \upsilon_0 = 0.0045$.
Since none of its components is assumed to fluctuate in time, the noise component is simply $\pmb{x}_s = 0$. It follows from \eqnref{eq:VdP} that $f_s(\pmb{x}, u, t, \pmb{\xi})$ has three terms:
\begin{align}
    f_s(\pmb{x}, u, t, \pmb{\xi}) = \begin{pmatrix}
        f_{\nu}(\pmb{x}, u, t, \pmb{\xi}) \\
        f_{\omega}(\pmb{x}, u, t, \pmb{\xi}) \\
        f_{\upsilon}(\pmb{x}, u, t, \pmb{\xi}) 
    \end{pmatrix}
\end{align}
where 
\begin{align}
    &f_{\nu}(\pmb{x}, u, t, \pmb{\xi}) = - p \, \omega(t) \\
    &f_{\omega}(\pmb{x}, u, t, \pmb{\xi}) = \frac{k}{m} \nu(t) + 2 \frac{c}{m} (1-\upsilon(t)) \, \omega(t)  \\
    &f_{\upsilon}(\pmb{x}, u, t, \pmb{\xi}) = \frac{|\nu(t)|-\nu(t)}{2 T} - \frac{\upsilon(t)}{T} 
\end{align}

%%%%%%%%%%%%%%%%%%%%%%%%%%%%%%%%%%%%%%%%%%%%%%%%%%%%%%%%%%%%%%%%%%%%%%%%%%%%%%%%%%%%%%%%%%%%%%%%%%%%
\appendixsection{Experimental parameters} 
\label{sec:exp_param}
\begin{figure}[t!]
    \centering
    \includegraphics[width=\linewidth]{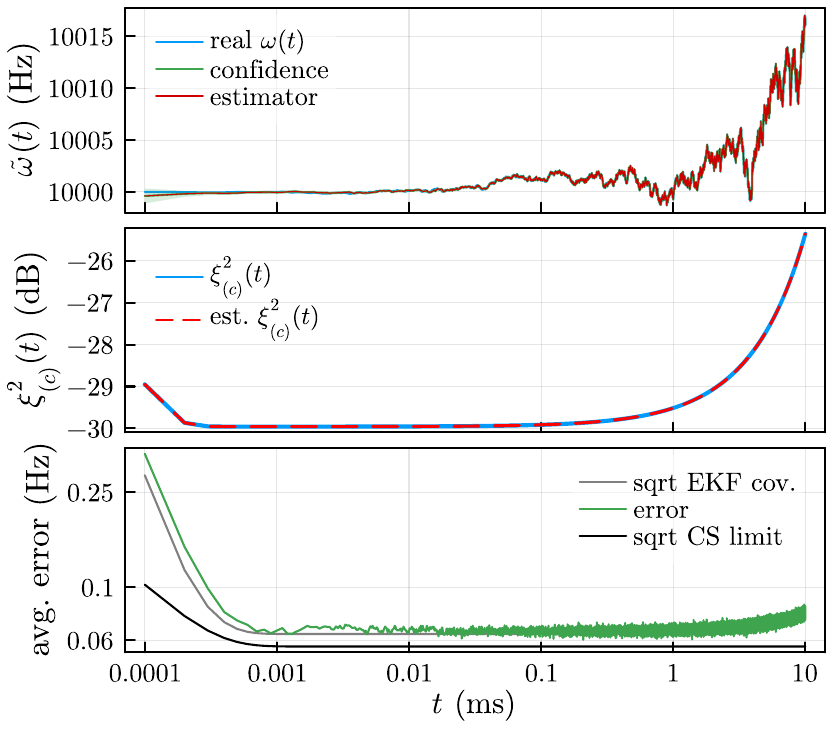}
    \caption{\textbf{Quantum-enhanced tracking of a fluctuating magnetic field with a higher measurement strength.} In the top plot, the OUP field (blue) is accurately tracked by its EKF estimate (red), remaining within within the error bounds of $\pm 2 \sqrt{\EE{\Delta^2 \est{\omega}(t)}}$, which are so small compared to the fluctuating field as to be nearly imperceptible. The middle plot shows the conditional spin squeezing (blue) generated by higher measurement strength of $M = \SI{1}{\milli\hertz} \gg \kcoll = \SI{1}{\nano\hertz}$, and its real-time estimation by the EKF (dashed red). In the bottom plot, the error in estimating $\omega(t)$ (green) achieves a sensitivity of $\sim 0.066 \trm{Hz}$ that matches the square-root of EKF covariance (dashed, yellow). The stronger measurement significantly reduces the error; however, the quantum limit set by the dephasing noise (black) at around $0.056 \trm{Hz}$ is not perfectly reached. An further increase in the measurement strength $M$ would yield an error closer to the optimal limit~\cite{Amoros-Binefa2024}. The bottom two plots have been obtained by averaging over 1000 field+atom trajectories. }
    \label{fig:largerM}
\end{figure}

The equation for the photocurrent in the main text, \eqnref{eq:photocurrent}, should be compared to Equation 18 from \refcite{Kong2020}. To ensure that the Wiener differential has a variance of $\dt$, as in \eqnref{eq:photocurrent}, the Eq.~(18) in \refcite{Kong2020} should be normalized by $\sqrt{\eta q_e^2 \dot{N}}$. Using the notation of this manuscript, the measurement equation of \refcite{Kong2020} can then be expressed as,
\begin{equation}
    I(t) \dt = \eta \sqrt{g^2 \dot{N}} \brktc{\Jy(t)} \dt + \sqrt{\eta} \, \dW,
    \label{eq:I(t)}
\end{equation}
where $\EE{\dW^2} = \dt$, and we use $\Jy$ instead of $\Jz$ to account for the different experimental geometry. By direct comparison to \eqnref{eq:photocurrent} we can identify the measurement strength $M$ as:
\begin{equation} \label{eq:def_M}
    M = \frac{g^2 \dot{N}}{4},
\end{equation}
where $\dot{N}$ is the photon flux given by:
\begin{equation}
    \dot{N} = \frac{P}{2\pi \hbar \, \nu}
\end{equation}
with $P$ being the probe power, ranging between \SI{0.5}{\milli\watt} and \SI{2}{\milli\watt}, and $\nu$ the frequency of the probe light, which is detuned $\Delta \nu = \nu_{\trm{D}_1\trm{Rb}} - \nu$ from the Rb $\trm{D}_1$ transition $\nu_{\trm{D}_1\trm{Rb}} = c/\lambda_{\trm{D}_1\trm{Rb}}$ of \SI{794.8}{\nano\meter}. The coupling constant $g$ in \eqnref{eq:def_M} is defined in \refcite{Kong2020} as:
\begin{equation}
    g \approx \frac{c \, r_\trm{e} f_\trm{osc}}{A_\trm{eff}} \frac{1}{\Delta \nu}
\end{equation}
where $c$ is the speed of light, $r_\trm{e} = \SI{2.82e-13}{\centi\meter}$ the classical electron radius, $f_\trm{osc} = 0.34$ the oscillator strength of the $\trm{D}_1$ transition in Rb and $A_\trm{eff} = \SI{0.0503}{\centi\meter^2}$ the effective beam area. Consequently, the value of $M$ will approximately range between $10^{-10}\SI{}{\hertz}$ and $10^{-8}\SI{}{\hertz}$ depending on the  probe power $P$ and optical detuning $\Delta \nu$, which when off-resonance can range between $\Delta\nu \approx \SIrange{24}{64}{\giga\hertz}$. 

Physically, one should interpret the parameter $M$ as an effective ratio between the light-atom interaction to the photon shot-noise in the detection process \eref{eq:I(t)}. However, the timescale at which the quantum backaction occurs due to continuous measurement is then given by $M'=M\,N$. This is because the quantum model \eref{eq:SME} yields a decay of the corresponding operator variance with an effective rate $1/M'$, as shown explicitly in absence of decoherence~\cite{Geremia2003} and with collective noise~\cite{Amoros-Binefa2021}, and consistently verified both here, recall \figref{fig:oup_sq_estimation}, and experimentally~\cite{Kong2020}.

%%%%%%%%%%%%%%%%%%%%%%%%%%%%%%%%%%%%%%%%%%%%%%%%%%%%%%%%%%%%%%%%%%%%%%%%%%%%%%%%%%%%%%%%%%%%%%
\appendixsection{EKF equations}
The form of the EKF equations will depend on the state space model, which changes depending on the form of the signal we aim to detect. Thus, the EKF equations need to be separately specified for the OUP and the VdP. 

%%%%%%%%%%%%%%%%%%%%%%%%%%%%%%%%%%%%%%%%%%%%%%%
\appendixsubsection{EKF for an OUP}
In particular, for an OUP like the one in \eqnref{eq:oup}, with $\EE{\dW_\omega^2} = q_\omega$, the Jacobians defining the Riccati equation in \eqnref{eq:Sigma_dyn} are:
\begin{widetext}
\begin{align}
    &F = \nabla_{\pmb{x}} \pmb{f} |_{(\pmb{\Tilde{x}},u,0)} = \nonumber \\
    &\left(
     \scalemath{0.75}{
    \begin{array}{ccccccc}
     -\left(\frac{M}{2} \!+\! \kloc \!+\! \frac{\kcoll}{2}\right) & -(\omega(t) \!+\! u) & 0 & 0 & 0 & 0 & -\estY \\ 
    (\omega(t) \!+\! u) & -\left(\frac{\kcoll + 2\kloc}{2}\right) & 0 & 0 & 0 & 0 & \estX \\ 
    0 & 2\kcoll\estY & -(M \!+\! 2\kloc \!+\! \kcoll) & \kcoll & M & -\left(8\eta M N \estCxy \!+\! 2(\omega(t) \!+\! u)\right) & -2\estCxy \\ 
    2\kcoll\estX & 0 & \kcoll & -(\kcoll \!+\! 2\kloc \!+\! 8\eta M N \estVy) & 0 & 2(\omega(t) \!+\! u) & 2\estCxy \\
    2M\estX & 0 & M & 0 & -M & 0 & 0 \\
    -\kcoll\estY & -\kcoll\estX & (\omega(t) \!+\! u) & -\left((\omega(t) \!+\! u) \!+\! 4\eta M N \estCxy\right) & 0 & -\left(\frac{M}{2} \!+\! 2\kcoll \!+\! 2\kloc \!+\! 4\eta M N \estVy\right) & \estVx \!-\! \estVy \\
    0 & 0 & 0 & 0 & 0 & 0 & -\chi_K
    \end{array}
    }
  \right) ,
\end{align}
\begin{align}
    G = \nabla_{\pmb{\xi}} \pmb{f}|_{\pmb{\Tilde{x}}} = 
    \begin{pmatrix} 
    2\sqrt{\eta M N} \estCxy & 0 \;\; \\ 
    2\sqrt{\eta M N} \estVy & 0 \;\; \\ 
    0 & 0 \;\; \\ 
    0 & 0 \;\; \\ 
    0 & 0 \;\; \\ 
    0 & 0 \;\; \\ 
    0 & 1 \;\;
    \end{pmatrix}, 
\end{align}
\begin{align}
    H = \nabla_{\pmb{x}} \pmb{h} &= \nabla_{\pmb{x}} \left(2\eta \sqrt{M N} \, \brktc{\Y} + \sqrt{\eta} \, \xi \right) = 2\eta \sqrt{M N} \begin{pmatrix} 0 & 1 & 0 & 0 & 0 & 0 & 0 \end{pmatrix},
\end{align}
\end{widetext}
of which $F$ and $G$ depending on the estimates and thus, having to be evaluated at each time-step by the estimator at that time $t$. Additionally, the matrices $Q \coloneqq \E{\pmb{\xi} \, \pmb{\xi}^\TT}{} =\Diag{1,q_K}$, $R \coloneqq \E{\zeta^2}{} = \eta$ and $S \coloneqq \E{\pmb{\xi} \zeta}{}= (\sqrt{\eta},\, 0)^\TT$ that appear in the Riccati equation (and in the Kalman gain $K$) correspond to the covariance and correlation matrices of the noise vectors and, importantly, are predetermined. Note that $F$ and $Q$ do not depend on $\chi$ or $q_\omega$ but we parametrize them with $\chi_K$ and $q_K$, to emphasize that these are KF parameters that might not exactly match the ones of the signal, $\chi$ and $q_\omega$, when we do not have access to that knowledge. 

%%%%%%%%%%%%%%%%%%%%%%%%%%%%%%%%%%%%%%%%%%%%%%%
\appendixsubsection{EKF for VdP}
\label{app:EKF_VdP}
When using a filtered Van der Pol oscillator to model a MCG-like signal, we assume that the second component of \eqnref{eq:VdP}, i.e. $\omega(t)$, is the one that couples to the atomic sensor. Nevertheless, we must consider the full system of \eqnref{eq:VdP} to derive the Jacobian and covariance matrices. In particular, 
\begin{widetext}
\begin{align}
    &\!\!\!\!\!\!F = \nabla_{\pmb{x}} \pmb{f} |_{(\pmb{\Tilde{x}},u,0)} = \nonumber \\
    &\!\!\!\!\!\!\!\!\!\!
    \left(
     \scalemath{0.7}{
    \begin{array}{ccccccccc}
     -\left(\frac{M}{2} \!+\! \kloc \!+\! \frac{\kcoll}{2}\right) & -(\omega(t) \!+\! u) & 0 & 0 & 0 & 0 & 0 & -\estY & 0 \\ 
    (\omega(t) \!+\! u) & -\left(\frac{\kcoll+2\kloc}{2}\right) & 0 & 0 & 0 & 0 & 0 & \estX & 0 \\ 
    0 & 2\kcoll\estY & -(M \!+\! 2\kloc \!+\! \kcoll) & \kcoll & M & -\left(8\eta M N \estCxy \!+\! 2(\omega(t) \!+\! u)\right) & 0 & -2\estCxy & 0 \\ 
    2\kcoll\estX & 0 & \kcoll & -(\kcoll \!+\! 2\kloc \!+\! 8\eta M N \estVy) & 0 & 2(\omega(t) \!+\! u) & 0 & 2\estCxy & 0 \\
    2M\estX & 0 & M & 0 & -M & 0 & 0 & 0 & 0 \\
    -\kcoll\estY & -\kcoll\estX & (\omega(t) \!+\! u) & -\left((\omega(t) \!+\! u) + 4\eta M N \estCxy\right) & 0 & -\left(\frac{M}{2} \!+\! 2\kcoll \!+\! 2\kloc \!+\! 4\eta M N \estVy\right) & 0 & \estVx \!-\! \estVy & 0 \\
    0 & 0 & 0 & 0 & 0 & 0 & 0 & -p_K & 0 \\
    0 & 0 & 0 & 0 & 0 & 0 & k_K/m_K & \frac{2c_K(1-\estVz)}{m_K} & -2c_K\estVy/m_K \\
    0 & 0 & 0 & 0 & 0 & 0 & -\frac{1+\estVx/|\estVx|}{2T_K} & 0 & -\frac{1}{T_K}
    \end{array}
    }
  \right) 
\end{align}
\begin{align}
    G = \nabla_{\pmb{\xi}} \pmb{f}|_{\pmb{\Tilde{x}}} = 
    \begin{pmatrix} 
    2\sqrt{\eta M} \estCxy & 0 \;\; \\ 
    2\sqrt{\eta M} \estVy & 0 \;\; \\ 
    0 & 0 \;\; \\ 
    0 & 0 \;\; \\ 
    0 & 0 \;\; \\ 
    0 & 0 \;\; \\ 
    0 & 0 \;\; \\ 
    0 & 1 \;\; \\
    0 & 0 \;\; 
    \end{pmatrix}, 
\end{align}
\begin{align}
    H = \nabla_{\pmb{x}} \pmb{h} &= \nabla_{\pmb{x}} \left(2\eta \sqrt{M} \, \brktc{\Jy} + \sqrt{\eta} \, \xi \right) = 2\eta \sqrt{M} \begin{pmatrix} 0 & 1 & 0 & 0 & 0 & 0 & 0 & 0 & 0 \end{pmatrix},
\end{align}
\end{widetext}
where the covariance and correlation matrices between the state noise and the measurement noise are $Q \coloneqq \E{\pmb{\xi} \, \pmb{\xi}^\TT}{} =\Diag{1,q_K}$, $R \coloneqq \E{\zeta^2}{} = \eta$ and $S \coloneqq \E{\pmb{\xi} \zeta}{}= (\sqrt{\eta},\, 0)^\TT$. The initial conditions are set to $\est{\pmb{x}}_0 = \left(\frac{\sqrt{N}}{2}, 0, 0, \frac{1}{4}, \frac{1}{4}, 0, \est{\nu}_0, \est{\omega}_0, \est{\upsilon}_0 \right)$ with $\est{\nu}_0 = \est{\omega}_0 = \est{\upsilon}_0 = 3.0045$.

%%%%%%%%%%%%%%%%%%%%%%%%%%%%%%%%%%%%%%%%%%%%%%%%%%%%%%%%%%%%%%%%%%%%%%%%%%%%%%%%%%%%%%%%%%%%%%
\appendixsection{Quantum precision limits in noisy systems} 
\label{ap:CSbound}
\appendixsubsection{General representation of a state under continuous measurement and measurement-based feedback}
Consider a generic map $\Phi_\Dt(\pmb{y}_k,\omega_k)$ acting on a state for a duration $\Dt$ according to all previous measurement records $\pmb{y}_k = \{y_0,y_1,\dots,y_k\}$, as well as a parameter $\omega_k$, the $k$th element of a time-discretized frequency trajectory $\pmb{\omega}_k = \{\omega_0,\omega_1,\dots,\omega_k\}$. These maps, which are applied at each time step, are interspersed with measurement POVMs $\measE_{y_k}$ such that $\sum_{k} \measE_{y_k}^\dagger \measE_{y_k}=\I$ and represent the discretised continuous measurement with outcome $y_k$. Hence, the state at time $k\Dt$, conditional on the measurement record  $\pmb{y}_k$, will read as
\begin{align} \label{eq:cm_discr} 
    &\rho(t|\pmb{y}_k) = \rhoc(k \Dt)  \\
    &\quad= \frac{\Phi_{k}\!\!\left[\measE_{y_{k}}    \dots \Phi_{1}\!\!\left[\measE_{y_{1}}\Phi_0\!\left[\measE_{y_{0}} \, \rho_{0} \measE_{y_{0}}^\dagger \right]\!\measE_{y_{1}}^{\dagger}\right] \dots \measE_{y_{k}}^{\dagger} \right]}{p(\pmb{y}_k|\pmb{\omega}_k)}, \nonumber 
\end{align}
where $\rhoc(t) \equiv \rho(t|\pmb{y}_k)$ refers to the conditional state at time $t$ and for convenience $\Phi_k \coloneqq \Phi_\Dt(\pmb{y}_k,\omega_k)$. The conditional probability in \eqnref{eq:cm_discr} of measuring outputs $\pmb{y}_k = \{y_0,y_1,\dots,y_k\}$ given field inputs $\pmb{\omega}_k = \{\omega_0,\omega_1,\dots,\omega_k\}$ is 
\begin{align} \label{eq:condit_prob}
    &p(\pmb{y}_k|\pmb{\omega}_k) =  \\
    &\quad=\trace{\!\Phi_{k}\!\!\left[\measE_{y_{k}}   \dots \Phi_{1}\!\!\left[\measE_{y_{1}}\Phi_0\!\left[\measE_{y_{0}}\rho_{0}\measE_{y_{0}}^\dagger\right]\!\measE_{y_{1}}^{\dagger}\right] \! \dots \measE_{y_{k}}^{\dagger} \!\right]}. \nonumber
\end{align}
As discussed in~\cite{Amoros-Binefa2024}, the POVMs $\measE_{y_k}$ for a homodyne-like continuous-measurement are the Kraus operators $\measE_{y_k} = \braketop{y_k}{\UnitOp_\Dt}{0}$ with the unitary governing the interaction between the system and the probe given by $\UnitOp_\Dt = \exp{\left\{-\sqrt{M \Dt} \left(\!  \LinOp  \otimes \bcreat_k  - \LinOp^\dagger \otimes \banihil_k \right) \right\} }$, where $\LinOp$ are the system's operators and $\banihil_k$ are the discretized modes of the probe~\cite{Albarelli2024}. Additionally, if we now consider $\Phi_{k}\!\left[ \; \cdot \; \right]$ to represent the most general form of measurement-based feedback, where the whole history of measurement results $\pmb{y}_{k}$ affects the Lindbladian governing the evolution of the state as
\begin{equation}
    \Phi_{k} [\,\cdot\,] = \ee^{\Lin_{\pmb{y}_k} \Dt} [\,\cdot\,] = [\,\cdot\,] + \Lin_{\pmb{y}_k} [\,\cdot\,] \Dt + \mathcal{O}(\Dt^2),
\end{equation}
then, the SME describing the state evolution can be written as
\begin{align}
    \dd\rhoc=\Lin_{\pmb{y}_{\leq t}} \rhoc \dt+M \D[\LinOp]\rhoc \dt+\sqrt{M}\H[\LinOp]\rhoc \dW,
\end{align}
where the measurement-induced nonlinear superoperator is $\H[ \GenOp] \rho \coloneqq  \GenOp \rho + \rho  \GenOp^\dagger - \trace{(\GenOp+\GenOp^\dagger)\rho}\rho$. 

It should be noted that the Lindbladian's dependency is exclusively on the measurement outcomes $\pmb{y}{\leq t}$, rather than on their rate of change, $I(t) = \frac{\dd y_{t}}{\dt} = \lim_{\Dt \rightarrow 0} \frac{\delta y_{t}}{\Dt}$.  However, if the Lindbladian turns out to depend on the derivative of the photocurrent, then, the framework outlined in \eqnref{eq:cm_discr} remains valid, but to replicate the evolution described in the references~\cite{Wiseman1994_feedback,Wiseman1994_squeezing_feedback}, a slightly different derivation must be followed~\cite{Amoros-Binefa2024}.

In cases where feedback only influences the unitary component of the Lindbladian:
\begin{align}
    \Lin_{\pmb{y}_{k}} [\,\cdot\,]
    =&-i (\omega(t)+ u(t|\pmb{y}_{k})) [\Jz,\,\cdot\,] \nonumber \\
    & + \frac{\kloc}{2} \sum_{j=1}^N \D[\sz^{(j)}] \,\cdot\, \, + \kcoll \D[\Jz]\,\cdot\,, 
\end{align}
we obtain a SME consistent with the sensor model described earlier in \eqnref{eq:SME}:
\begin{align}
    d\rhoc = &-i (\omega(t)+ u(t|\pmb{y}_{k})) [\Jz,\rhoc] \dt \nonumber \\
    &+ \frac{\kloc}{2} \sum_{j=1}^N \D[\sz^{(j)}] \; \rhoc \dt \, + \kcoll \D[\Jz]\rhoc \dt  \nonumber \\
    &+ M\D[\LinOp]\rhoc \dt + \sqrt{ M} \H[\LinOp] \rhoc \dW. \label{eq:SME_feedback}
\end{align} 

Accordingly, the transformation $ \Phi_{k} [\,\cdot\,]$, corresponding to the non-measurement components in \eqnref{eq:SME_feedback}, is defined as the sequential application of the following operations:
\begin{equation} \label{eq:concatenated_maps}
\Phi_{k} = \Omega_k \circ \Lambda_{\omega,k} \circ \FF_{\pmb{y}k},
\end{equation}
where $\FF_{\pmb{y}k}$ accounts for the feedback, $\Lambda_{\omega,k}$ represents the unitary evolution (parametrized with $\omega$) and local dissipative effects (with dissipation rate $\kloc$), and $\Omega_k$ handles the collective decoherence (with strength $\kcoll$). While these operators act on $\Jz$ and are therefore commutative, a Suzuki-Trotter expansion to the first order in $\Dt$ would still be applicable for non-commutative operations, enabling them to be handled sequentially as necessary. Then, \eqnref{eq:cm_discr} becomes

\begin{align}
    \!\!&\;
    \rhoc\,[k] = \\
    &=\! \frac{\Omega_k\!\!\left[\Lambda_{\omega,k} \!\!\left[ \FF_{\pmb{y}_k}\!\!\left[\!\measE_{y_{k}} \! \dots \Omega_0\!\!\left[\Lambda_{\omega,0} \!\!\left[ \FF_{\pmb{y}_0}\!\!\left[\measE_{y_{0}}\rho_{0}\measE_{y_{0}}^\dagger\!\right]\!\right]\!\right]\!\! \dots  \! \measE_{y_{k}}^{\dagger} \!\right]\!\right]\!\right]}{p(\pmb{y}_k|\pmb{\omega}_k)},  \nonumber
\end{align}
such that 
\begin{align} \label{eq:discretized_likelihood}.
    \!\!&\;
    p(\pmb{y}_k|\pmb{\omega}_k) \!=\! \\
    &=\!\trace{\!\Omega_k\!\!\left[\Lambda_{\omega,k} \!\!\left[ \FF_{\pmb{y}_k}\!\!\left[\!\measE_{y_{k}}   \! \dots \Omega_0\!\!\left[\Lambda_{\omega,0} \!\!\left[ \FF_{\pmb{y}_0}\!\!\left[\!\measE_{y_{0}}\rho_{0}\measE_{y_{0}}^\dagger\!\right]\!\right]\!\right]\! \!\measE_{y_{k}}^{\dagger} \!\right]\!\right]\!\right]\!} \nonumber 
\end{align}

%%%%%%%%%%%%%%%%%%%%%%%%%%%%%%%%%%%%%%%%%%%%%%%
\appendixsubsection{Convex decomposition of the likelihood}
Following the same steps as in~\cite{Amoros-Binefa2024}, our goal is to find a convex decomposition of the joint noisy map~$\Omega\left[\Lambda_{\omega}\left[ \; \cdot \; \right]\right]$ which also accounts for the $\omega$-encoding, in order to decompose the discretized likelihood \eref{eq:discretized_likelihood} as
\begin{align} \label{eq:likelihood_convex_decomp}
    p(\pmb{y}_k|\pmb{\omega}_k) = \int \D\pmb{\bigzeta}_k \; q(\pmb{\bigzeta}_k|\pmb{\omega}_k) \; p(\pmb{y}_k|\pmb{\bigzeta}_k)
\end{align}
where $\pmb{\bigzeta}_k = \{\pmb{\zeta}_0,\pmb{\zeta}_1,\dots,\pmb{\zeta}_k\}$ is a sequence of sets, each containing $N$ auxiliary frequency-like random variables. For instance, $\pmb{\zeta}_\ell = \{\zeta^{\,(1)}_\ell,\zeta^{\,(2)}_\ell, \dots, \zeta^{\,(N)}_\ell\}$ indicates that within the $\ell$th step, the first probe undergoes the Larmor precession for $\Dt$ with frequency $\zeta^{\,(1)}$, the second probe with $\zeta^{\,(2)}$ etc. 

While $q(\pmb{\bigzeta}_k|\pmb{\omega}_k)$ represents the mixing distribution that crucially contains all the $\pmb{\omega}_k$-dependence, $p(\pmb{y}_k|\pmb{\bigzeta}_k)$ in \eqnref{eq:likelihood_convex_decomp} can be interpreted as a (fictitious) likelihood of obtaining the measurement record $\{\pmb{y}_j\}_{j=1}^k$, where the discretised measurements and feedback are interspersed by unitary maps undergoing frequency encoding as specified by the sequence $\pmb{\bigzeta}_k$, i.e.: 
\begin{align}
    \!\!&\; \label{eq:likelihood_Zk_Uk}
    p(\pmb{y}_k|\pmb{\bigzeta}_k) = \\ &=\trace{\mathcal{U}_{\pmb{\zeta}_k}\!\left[\FF_{\pmb{y}_{k}}\!\left[ \measE_{y_{k}} \! \dots \,\mathcal{U}_{\pmb{\zeta}_0}\!\left[\FF_{\pmb{y}_0}\!\left[\measE_{y_{0}}\rho_{0}\measE_{y_{0}}^\dagger\right]\right]\!\dots\!\measE_{y_{k}}^{\dagger}\! \right]\right]\!}\!. \nonumber 
\end{align} 

To find such a convex decomposition for the overall map $\Omega\left[\Lambda_{\omega}\!\left[ \; \cdot \; \right]\right]$ and thus be able to write the discretized likelihood $p(\pmb{y}_k|\pmb{\bigzeta}_k) $ as \eqnref{eq:likelihood_Zk_Uk}, we will express it as a mixture of unitaries by decomposing separately the collective map $\Omega[\cdot]$ that acts on all the probes, and the local map $\Lambda_{\omega}[\cdot]$ that exhibits a tensor product structure with local decoherence and unitary evolution acting independently on each probe~\cite{Amoros-Binefa2024}. Namely, 
\begin{align} \label{eq:unitary_form_joint_map}
        \Omega[\Lambda_\omega[\; \cdot \; ]] &=  \frac{1}{\sqrt{2\pi \Vq}}\int \! \! \D\pmb{\zeta} \; \fzeta \ee^{- \frac{(\avgzeta - \omega)^2}{2\Vq}} \Unitary_{\pmb{\zeta}}[ \; \cdot \; ] 
\end{align}
where $\avgzeta \coloneqq \frac{1}{N} \sum_{i = 1}^N \zetai$, the ``effective'' variance is
\begin{equation} \label{eq:defVq}
    \Vq = \Vcoll + \Vloc/N = \frac{\kcoll + 2 \kloc/N}{\Dt} = \frac{\kq}{\Dt},
\end{equation}
with $\kq\coloneqq\kcoll + 2 \kloc/N$, and
\begin{equation}
    \fzeta = \sqrt{\frac{1}{N (2\pi \Vloc)^{N-1}}} \ee^{\left\{- \frac{1}{2\Vloc} \left(\sum_{i=1}^N  (\zetai)^2 -  N \avgzeta^2 \right) \right\}}.
\end{equation}

%%%%%%%%%%%%%%%%%%%%%%%%%%%%%%%%%%%%%%%%%%%%%%%
\appendixsubsection{Upper bound on the Fisher Information}
The main goal behind our proof is to bypass the calculation of the Fisher Information $\Fisher[p(\pmb{y}_{k}|\omega_k)]$, defined as
\begin{align}
    \Fisher[p(\pmb{y}_{k}|\omega_k)] &= \E{ \left(\partial_{\omega_k} \log{p(\pmb{y}_{k}|\omega_k)} \right)^2}{p(\pmb{y}_{k}|\omega_k)} \label{eq:other_def_FI} \\
    &= \E{-\partial^2_{\omega_k} \log{p(\pmb{y}_{k}|\omega_k)}}{p(\pmb{y}_{k}|\omega_k)}, \label{eq:best_def_FI}
\end{align}
in the Bayesian Cram\'{e}r-Rao Bound (BCRB) by upper-bounding it. However, the likelihood $p(\pmb{y}_k|\pmb{\omega}_k)$ decomposed in \eqnref{eq:likelihood_convex_decomp} is not exactly the conditional distribution appearing in the BCRB. In particular, the BCRB, which is defined as the inverse of the Bayesian Information $J_B$,
\begin{align} \label{eq:BCRB}
    \EE{\Delta^2\est{\omega}_k} &\geq (J_B)^{-1} \nonumber \\
    &\geq \frac{1}{\Fisher[p(\omega_k)] + \int \dd \omega_k \, p(\omega_k) \Fisher[p(\pmb{y}_{k}|\omega_k)]},
\end{align}
depends on both the prior knowledge about the frequency $\omega_k$ at time $k\Dt$ and $p(\pmb{y}_{k}|\omega_k)$, and the probability of recording a measurement trajectory $\pmb{y}_{k}$ given that the frequency at time $t$ is $\omega_k$. Thus, we must find a way to relate the likelihood  $p(\pmb{y}_{k}|\omega_k)$ required to compute the BCRB to the likelihood $p(\pmb{y}_k|\pmb{\omega}_k)$ appearing in \eqnref{eq:likelihood_convex_decomp}, i.e. the likelihood of detecting a measurement record $\pmb{y}_k$ given that the Larmor frequency has followed a trajectory $\pmb{\omega}_k = \{\omega_0,\omega_1,\dots,\omega_k\}$. To do so, we simply use the Bayes' rule and rewrite
\begin{align} \label{eq:p(y|w)_bayes_rewritten}
    p(\pmb{y}_{k}|\omega_k) &= \frac{p(\pmb{y}_{k},\omega_k)}{p(\omega_k)} = \frac{1}{p(\omega_k)} \!\! \int \!\! \dd \pmb{\omega}_{k\shortminus1} \, p(\pmb{y}_k,\pmb{\omega}_k) \nonumber \\
    &= \frac{1}{p(\omega_k)} \!\! \int \!\! \dd \pmb{\omega}_{k\shortminus1} \, p(\pmb{\omega}_k) p(\pmb{y}_k|\pmb{\omega}_k).  
\end{align}
Thus, now we can apply  \eqnref{eq:likelihood_convex_decomp} to \eqnref{eq:p(y|w)_bayes_rewritten} and reveals a decomposition similar to \eqnref{eq:likelihood_convex_decomp} but for $p(\pmb{y}_{k}|\omega_k)$. Specifically,
\begin{align}
    p(\pmb{y}_{k}|\omega_k) \!&= \!\! \int \!\! \D\pmb{\bigzeta}_k p(\pmb{y}_k|\pmb{\bigzeta}_k) \! \left[ \frac{1}{p(\omega_k)} \!\! \int \!\! \dd \pmb{\omega}_{k\shortminus1} p(\pmb{\omega}_k) q(\pmb{\bigzeta}_k|\pmb{\omega}_k) \! \right] \nonumber \\
    &= \!\! \int \!\! \D\pmb{\bigzeta}_k  p(\pmb{y}_k|\pmb{\bigzeta}_k)  \BigP_{\omega_k}(\pmb{\bigzeta}_k) \nonumber \\ 
    &= \! \mapS_{\pmb{\bigzeta}_{k}\to\pmb{y}_{k}}\!\left[\BigP_{\omega_k}(\pmb{\bigzeta}_{k})\right]
\end{align}
where we identify $\mapS_{\pmb{\bigzeta}_{k}\to\,\pmb{y}_{k}}[\; \bigcdot \; ]=\int\! d\vec{\bigzeta}_{k} \,  p(\pmb{y}_{k}|\,\pmb{\bigzeta}_{k})\; \bigcdot \;$ as a stochastic map independent of the parameter $\omega_k$, and the probability distribution $\BigP_{\omega_k}(\pmb{\bigzeta}_{k})$ as
\begin{equation} \label{eq:def_BigP}
    \BigP_{\omega_k}(\pmb{\bigzeta}_{k}) = \frac{1}{p(\omega_k)} \!\! \int \!\! \dd \pmb{\omega}_{k\shortminus1} \, p(\pmb{\omega}_k) q(\pmb{\bigzeta}_k|\pmb{\omega}_k), 
\end{equation}
which contains the information on $\omega_k$ within $q(\pmb{\bigzeta}_k|\pmb{\omega}_k)$. 
As the Fisher Information is generally contractive (monotonic) under the action of stochastic maps, we can now upper-bound $\Fisher[p(\pmb{y}_{k}|\omega_k)]$ as
\begin{align}
    \Fisher[p(\pmb{y}_{k}|\omega_k)] = \Fisher[\mapS_{\pmb{\bigzeta}_{k}\to\pmb{y}_{k}}\!\left[\BigP_{\omega_k}(\pmb{\bigzeta}_{k})\right]] \leq \Fisher[\BigP_{\omega_k}(\pmb{\bigzeta}_{k})].
\end{align}

Thus, the problem of lower-bounding the BCRB in \eqnref{eq:BCRB} now reduces to evaluating the Fisher Information of $\BigP_{\omega_k}(\pmb{\bigzeta}_{k})$.

\appendixsubsection{Analytical form of $\BigP_{\omega_k}(\pmb{\bigzeta}_{k})$}
The probability distribution outlined in \eqnref{eq:def_BigP} consists of three different probability distributions: the marginal probability $p(\omega_k)$, the prior distribution $p(\pmb{\omega}_k)$ and the classically-simulated likelihood or mixing distribution $q(\pmb{\bigzeta}_k|\pmb{\omega}_k)$. To derive an analytical expression for \eqnref{eq:def_BigP}, we first need to elaborate on the exact forms of each probability component. 
\appendixsubsubsection{Prior contribution}
Tracking is a sensing task where we wish to monitor the evolution of a parameter, in this case a trajectory of frequencies $\pmb{\omega}_k = \{\omega_0,\omega_1,\dots,\omega_k\}$, with each element $\omega_k$ drawn from a probability distribution $p(\omega_k)$:
\begin{equation}
    p(\omega_k) = \int \dd \pmb{\omega}_{k\shortminus1} p(\pmb{\omega}_k).
\end{equation}

We choose $\omega_k$ to be the time-discretized version of the OU process $\omega(t)$ of \eqnref{eq:oup}. However, in the following proof, we omit the term $\bar{\omega}$ since it has no impact on the aMSE of $\est{\omega}$. Specifically, shifting a process by a deterministic value $\bar{\omega}$, i.e. $\nu(t) \coloneqq \omega(t) - \bar{\omega}$, preserves the aMSE: $\EE{\Delta^2 \est{\nu}(t)} = \EE{\Delta^2 \est{\omega}(t)}$. Hence, the effective process we consider is:
\begin{equation}
    \dd \omega(t) = - \chi \omega(t) \dt + \sqrt{q_\omega} \dW_\omega
\end{equation}
where $\chi > 0$ and $q_\omega > 0$ parametrize the decay and volatility of the process, and $\dW_\omega$ denotes the Wiener differential with mean $\EE{\dW_\omega} = 0$ and variance $\EE{\dW^2_\omega} = \dt$, then the probability of the process transitioning from $\omega_{k\shortminus1}$ at time $(k-1)\Dt$ to $\omega_{k}$ at $k\Dt$ is given by
\begin{equation} \label{eq:transition_prob_OUP}
    p(\omega_k|\omega_{k\shortminus1}) = \sqrt{\frac{1}{2\pi \Vp }} \ee^{ - \dfrac{(\omega_k - \omega_{k\shortminus1} \ee^{-\chi \delta t})^2}{2\Vp} }
\end{equation}
with variance 
\begin{equation} \label{eq:VarP}
    \Vp = \frac{q_\omega}{2\chi} (1-\ee^{-2\chi\Dt}).
\end{equation}
Since the Ornstein-Uhlenbeck process is a Markov process, the probability of the process $\omega(t)$ of following a discrete trajectory $\pmb{\omega}_k = \{ \omega_0,\omega_1,\dots,\omega_k\}$ is given by
\begin{equation} \label{eq:traj_p(omega_k)_OUP}
    p(\pmb{\omega}_k) = \prod_{i=1}^k p(\omega_i|\omega_{i-1}) p(\omega_0),
\end{equation}
where we assume $p(\omega_0)$ to be a Gaussian prior with mean zero and variance $\sigma_0^2$, i.e. $p(\omega_0) = \mathcal{N}(0,\sigma_0^2)$. From that, we can compute the probability of the frequency taking the value $\omega_k$ at time $k\Dt$, irrespective of the previous values of $\omega$:
\begin{align} \label{eq:p(omega_k)_OUP}
    p(\omega_k) = \frac{1}{\sqrt{2\pi \Vp^{(k)}}}\, \exp\!\left(-\frac{\omega_k^2}{2\Vp^{(k)}}\right),
\end{align}
with variance
\begin{equation}\label{eq:VarPk}
  \Vp^{(k)} = \sigma_0^2 \ee^{-2k\chi\Dt} + \frac{q_\omega}{2\chi}(1 - \ee^{-2k\chi\Dt}).
\end{equation}

%%%%%%%%%%%%%%%%%%%%%%%
\appendixsubsubsection{Classically-simulated contribution}
If now we substitute the classically-simulated form of the joint map $\Omega[\Lambda_\omega[\;\cdot\;]]$ as written in \eqnref{eq:unitary_form_joint_map} into  the likelihood $p(\pmb{y}_k|\pmb{\omega}_k)$ in \eqnref{eq:discretized_likelihood}, then we retrieve the desired decomposition of \eqnref{eq:likelihood_convex_decomp}. Namely,
\begin{align}
    \!\!&\;
    p(\pmb{y}_k|\pmb{\omega}_k) = \nonumber \\
    &=  \! \int \!\! \D\pmb{\bigzeta}_k \! \left[ \prod_{j=0}^k \fzetaindex{j} \frac{1}{\sqrt{2\pi\Vq}}\,\ee^{- \frac{(\avgzeta_j - \omega_j)^2}{2\Vq}} \right] \! p(\pmb{y}_k|\pmb{\bigzeta}_k) \nonumber \\
    &= \int \D\pmb{\bigzeta}_k \; q(\pmb{\bigzeta}_k|\pmb{\omega}_k) \; p(\pmb{y}_k|\pmb{\bigzeta}_k),
\end{align}
with $p(\pmb{y}_k|\pmb{\bigzeta}_k)$ being \eqnref{eq:likelihood_Zk_Uk}.
Thus, we identify $q(\pmb{\bigzeta}_k|\pmb{\omega}_k)$ as a product of distributions
\begin{align}
    q(\pmb{\bigzeta}_k|\pmb{\omega}_k) &= \prod_{j=0}^k q(\pmb{\zeta}_j|\omega_j) = \prod_{j=0}^k \fzetaindex{j} \frac{1}{\sqrt{2\pi\Vq}} \ee^{- \frac{(\avgzeta_j - \omega_j)^2}{2\Vq}} \nonumber \\
    &= f\,(\pmb{\bigzeta}_k) \mathcal{Q}(\pmb{\avgzeta}_k|\pmb{\omega}_k),
\end{align}
where $f\,(\pmb{\bigzeta}_k) = \prod_{j=0}^k \fzetaindex{j}$ and
\begin{align}
    \mathcal{Q}(\pmb{\avgzeta}_k|\pmb{\omega}_k) &= \prod_{j=0}^k \mathcal{Q}(\avgzeta_j|\omega_j) \nonumber \\ 
    &=\prod_{j=0}^k \frac{1}{\sqrt{2\pi\Vq}} \ee^{- \frac{(\avgzeta_j - \omega_j)^2}{2\Vq}}
\end{align}
is made up of a product of $k+1$ Gaussians $\mathcal{Q}(\avgzeta_j|\omega_j)$ with mean $\omega_j$ and variance $\Vq = \Vcoll + \Vloc/N$.

%%%%%%%%%%%%%%%%%%%%%%%%%%%
\appendixsubsubsection{Integrated form of $\BigP_{\omega_k}(\pmb{\bigzeta}_{k})$}
Once each contribution to $\BigP_{\omega_k}(\pmb{\bigzeta}_{k})$ has been established, we can bring them all together and rearrange its integral form \eqref{eq:def_BigP} as a set of nested integrals:
\begin{align} \label{eq:rearranging_BigP}
    &\BigP_{\omega_k}(\pmb{\bigzeta}_{k}) = \frac{f(\pmb{\bigzeta}_k)}{p(\omega_k)} \!\! \int \!\! \dd \pmb{\omega}_{k\shortminus1} \, p(\pmb{\omega}_k) \mathcal{Q}(\pmb{\avgzeta}_k|\pmb{\omega}_k)  \\
    &= \frac{f(\pmb{\bigzeta}_k)}{p(\omega_k)} \!\!\!\int\!\!\! \dd \pmb{\omega}_{k\shortminus1} \!\! \prod_{j=1}^k \! p(\omega_j|\omega_{j\shortminus1}) \mathcal{Q}(\avgzeta_j|\omega_j) \, p(\omega_0) \mathcal{Q}(\avgzeta_0|\omega_0) \nonumber \\
    &= \frac{f(\pmb{\bigzeta}_k)}{p(\omega_k)} \mathcal{Q}(\avgzeta_k|\omega_k) \!\!\int\!\! \dd \omega_{k\shortminus1} \, p(\omega_k|\omega_{k\shortminus1}) \mathcal{Q}(\avgzeta_{k\shortminus1}|\omega_{k\shortminus1}) \nonumber \\
    & \, \dots \!\!\int\!\! \dd \omega_1 \, p(\omega_2|\omega_1) \mathcal{Q}(\avgzeta_1|\omega_1) \!\!\int\!\! \dd \omega_0 \,  p(\omega_1|\omega_0) \mathcal{Q}(\avgzeta_0|\omega_0) p(\omega_0). \nonumber
\end{align}

Since all the functions inside the integrals are Gaussian, this set of nested integrals has a recursive solution, as given in~\cite{Amoros-Binefa2021}. In particular, if we identify the variances of the recursive relation in the Lemma 2 of~\cite{Amoros-Binefa2021}, $\Vp$ and $\Vq$, with \eqnref{eq:VarP} and $\Vq = \Vcoll + \Vloc/N$, respectively, we can then simply state that,
\begin{align}
    \BigP_{\omega_k}(\pmb{\bigzeta}_{k}) &= \frac{f\,(\pmb{\bigzeta}_k)}{p(\omega_k)} \mathcal{Q}(\avgzeta_k|\omega_k) \mathcal{P}_k(\omega_k)
\end{align}
where $\mathcal{P}_k(\omega_k)$ is a Gaussian distribution 
\begin{equation}
    \mathcal{P}_k(\omega_k) = C_k \ee^{-\frac{(\omega_k - \mu_k)^2}{2\mrm{V}_k}}
\end{equation}
and its variance follows the recursive relation
\begin{equation} \label{eq:variance_recursive_relation}
    \mrm{V}_k = \Vp + \frac{\Vq V_{k\shortminus1}}{\Vq + V_{k\shortminus1}}
\end{equation}
with initial value $V_0 = \sigma_0^2$.

%%%%%%%%%%%%%%%%%%%%%%%%%%%%%%%%%%%%%%%%%%%%%%%
\appendixsubsection{Fisher Information of $\BigP_{\omega_k}(\pmb{\bigzeta}_{k})$}
Now that we have a closed form for $\BigP_{\omega_k}(\pmb{\bigzeta}_{k})$, we can move to computing its Fisher Information using the form for the Fisher specified in \eqnref{eq:best_def_FI}. Namely,
\begin{align}
    &\Fisher{[\BigP_{\omega_k}(\pmb{\bigzeta}_{k})]} = \nonumber \\
    &= \int \dd \pmb{\bigzeta}_k \BigP_{\omega_k}(\pmb{\bigzeta}_{k}) \! \left[ \! -\partial^2_{\omega_k} \! \log{\!\left( \frac{f\,(\pmb{\bigzeta}_k)}{p(\omega_k)} \mathcal{Q}(\avgzeta_k|\omega_k) \mathcal{P}_k(\omega_k) \!\right)}\!\right] \nonumber \\
    & = \int \dd \pmb{\bigzeta}_k \BigP_{\omega_k}(\pmb{\bigzeta}_{k}) \left[-\partial^2_{\omega_k} \log{ f\,(\pmb{\bigzeta}_k) } \right] \label{eq:Fisher_of_f} \\
    & - \int \dd \pmb{\bigzeta}_k \BigP_{\omega_k}(\pmb{\bigzeta}_{k}) \left[-\partial^2_{\omega_k} \log{p(\omega_k) } \right]  \label{eq:Fisher_of_p} \\
    & + \int \dd \pmb{\bigzeta}_k \BigP_{\omega_k}(\pmb{\bigzeta}_{k}) \left[-\partial^2_{\omega_k} \log{ \mathcal{Q}(\avgzeta_k|\omega_k) } \right]  \label{eq:Fisher_of_curlyQ}\\
    & + \int \dd \pmb{\bigzeta}_k \BigP_{\omega_k}(\pmb{\bigzeta}_{k}) \left[-\partial^2_{\omega_k} \log{\mathcal{P}_k(\omega_k)  } \right]  \label{eq:Fisher_of_curlyP} \\
    &= -\frac{1}{\Vp^{(k)}} + \frac{1}{\Vq} + \frac{1}{\mrm{V}_k} \label{eq:Fisher_variances_form},
\end{align}
where to reach the final expression of \eref{eq:Fisher_variances_form} we have used the fact that $f\,(\pmb{\bigzeta}_k)$ in \eqnref{eq:Fisher_of_f} does not depend on $\omega_k$, and that in \eqnsref{eq:Fisher_of_p}{eq:Fisher_of_curlyP}, $-\partial_{\omega_k}^2 \log{\ee^{-\frac{(\omega_k - \mu)^2}{2 V}}} = V^{-1}$. We have a closed expression for both $\Vp^{(k)}$ and $\Vq$, but not for $\mrm{V}_k$. For that, we need to solve the recursive relation of \eqnref{eq:variance_recursive_relation}. Fortunately, such a form for $\mrm{V}_k$ exists, even though lengthy:
\begin{align}
    \mrm{V}_k = \frac{W_+ V_+^{\,k} + W_-V_-^{\,k}}{U_-V_-^{\,k} + U_+ V_+^{\,k}},
\end{align}
with terms $W_+$, $V_+$, $U_+$, $W_-$, $V_-$, and $U_-$ given by
\begin{align}
    W_+ &= 2\Vp\Vq + \sigma_0^2 \Vp + \sigma_0^2 \sqrt{\Vp(4\Vq + \Vp)} \\
    W_- &= -2\Vp\Vq - \sigma_0^2 \Vp + \sigma_0^2 \sqrt{\Vp(4\Vq + \Vp} \\
    U_+ &= -\Vp + 2\sigma_0^2 + \sqrt{\Vp(4\Vq + \Vp)} \\
    U_- &= \Vp - 2\sigma_0^2 + \sqrt{\Vp(4\Vq + \Vp)}\\ 
    V_+ &= 2\Vq + \Vp + \sqrt{\Vp(4\Vq + \Vp)}\\
    V_- &= 2\Vq + \Vp - \sqrt{\Vp(4\Vq + \Vp)}
\end{align}
where $\Vq$ is the variance given in \eqnref{eq:defVq} and $\Vp$ is defined in \eqnref{eq:VarP}.

%%%%%%%%%%%%%%%%%%%%%%%%%%%%%%%%%%%%%%%%%%%%%%%
\appendixsubsection{Continuous-time limit}
If now we take the continuous-time limit of $\Dt \rightarrow 0$, we can observe that the term $1/\Vq$ in \eqnref{eq:Fisher_variances_form} goes to zero. The other terms become,
\begin{align}
    \Vp(t) &= \lim_{\Dt \rightarrow 0} \Vp^{(k)} = \sigma_0^2 \ee^{-2\chi t} + \frac{q_\omega}{2\chi}(1 - \ee^{-2\chi t})
\end{align}
and
\begin{align}
    \mrm{V}_{\sigma_0}(t) &= \lim_{\Dt \rightarrow 0} V_{k}  \nonumber \\
    &= \frac{\sqrt{q_\omega \kq} \, \sigma_0^2 \cosh{\left(t\sqrt{\dfrac{q_\omega}{\kq}}\right)} + q_\omega \kq \sinh{\left(t\sqrt{\dfrac{q_\omega}{\kq}}\right)}}{\sqrt{q_\omega \kq} \cosh{\left(t\sqrt{\dfrac{q_\omega}{\kq}}\right)} + \sigma_0^2 \sinh{\left(t \sqrt{\dfrac{q_\omega}{\kq}}\right)}},
    \label{eq:V_CSlim}
\end{align}
where $\kq = \kcoll + 2\kloc/N$. Therefore, the BCRB \eref{eq:BCRB} can be now bounded as follows
\begin{align}
    \EE{\Delta^2\est{\omega}_k} 
    %&\geq \frac{1}{\Fisher[p(\omega_k)] + \int \dd \omega_k \, p(\omega_k) \Fisher[p(\pmb{y}_{k}|\omega_k)]} \nonumber \\
    &\geq \frac{1}{\Fisher[p(\omega_k)] + \int \dd \omega_k \, p(\omega_k) \Fisher[\BigP_{\omega_k}(\pmb{\bigzeta}_{k})]} \nonumber \\
    &= \dfrac{1}{\dfrac{1}{\Vp^{(k)}} - \dfrac{1}{\Vp^{(k)}} + \dfrac{1}{\mrm{V}_k}} = \mrm{V}_k,
\end{align}
where we have used the fact that $\Fisher[p(\omega_k)] = 1/\Vp^{(k)}$. 

Hence, in its most general form in the continuous time limit, the aMSE is lower-bounded by
\begin{align}
    &\EE{\Delta^2\est{\omega}(t)} \! \geq \mrm{V}_{\sigma_0}(t)  \\
    &=\!\dfrac{\sqrt{q_\omega \kq(N)} \, \sigma_0^2 \cosh{\left(\!t\sqrt{\dfrac{q_\omega}{\kq(N)}}\right)} \!+\! q_\omega \kq(N) \sinh{\left(\!t\sqrt{\dfrac{q_\omega}{\kq(N)}}\right)}}{\sqrt{q_\omega \kq(N)} \cosh{\left(\!t\sqrt{\dfrac{q_\omega}{\kq(N)}}\right)} \!+\! \sigma_0^2 \sinh{\left(\!t \sqrt{\dfrac{q_\omega}{\kq(N)}}\right)}} \nonumber
\end{align}
which in the limit $\sigma_0 \rightarrow \infty$ simplifies to
\begin{align} \label{eq:CSlimit_simplified}
    \!\!\!\EE{\Delta^2\est{\omega}(t)} &\geq \mrm{V}_{\infty}(t,N) \nonumber \\
    &= \sqrt{q_\omega \kq(N)} \coth{\left(t \sqrt{\frac{q_\omega}{\kq(N)}}\right)},
\end{align}
and matches the result of \refcite{Amoros-Binefa2021} if further $\kloc=0\,\implies\,\kq(N) = \kcoll$. If we then further take the limit of $q_\omega \rightarrow 0$, it then becomes
\begin{equation}
    \EE{\Delta^2\est{\omega}(t)} \geq \frac{\kq(N)}{t} = \frac{\kcoll}{t} + \frac{2\kloc}{N t},
\end{equation}
which now exhibits the standard quantum limit (SQL), as discussed in \refcite{Amoros-Binefa2024}.

\begin{figure}[t!]
    \includegraphics[width=\linewidth]{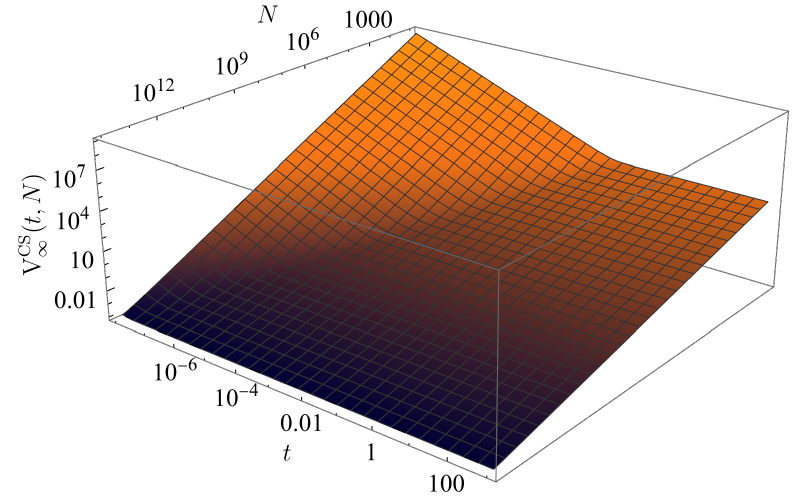}
    \caption{\textbf{Plot w.r.t.~$N$ and $t$ of the bound \eref{eq:CSlimit_simplified} on the aMSE imposed by dephasing and field fluctuations}. Log-log plot of the aMSE bound $\mrm{V}_\infty(t,N)$, showcasing the convex behaviour of the function over several orders of magnitude of $N$ and $t$. The colour gradient indicates the magnitude of the function, transitioning from high (bright) values to low (dark) values. The parameters used to generate this figure: $\sigma_0 = 10, \; q_\omega = \SI{1e6}{\radian^{2} \, \second^{-3}}, \; \kcoll = 0, \; \kloc = \SI{100}{\hertz}$.}
    \label{fig:2Dplot_CSlimit}
\end{figure}

%%%%%%%%%%%%%%%%%%%%%%%%%%%%%%%%%%%%%%%%%%%%%%%
\appendixsubsection{Accounting for fluctuations of the atomic number $N$}
In typical atomic magnetometry experiments, the number of atoms $N$ is not precisely known and changes from shot to shot. To model this, we assume that in each realization of the experiment, the actual atomic number is drawn from a Gaussian distribution with mean $\bar{N} = 10^{13}$ and standard deviation $\sigma = 10^{11}$~\cite{Kong2020}, i.e.:
\begin{equation}
    N \sim p(N) = \mathcal{N}(\bar{N},\sigma^2).
\end{equation}
While the system evolves with the specific drawn value of $N$, the estimation is performed assuming that $N = \bar{N}$, since the experimenter does not have access to the exact number of atoms in the ensemble. 
Consequently, the aMSE must be further averaged over the probability distribution of $N$, $p(N)$. 

As a result, the bound on the aMSE given in \eqnref{eq:CSlimit_simplified} becomes:
\begin{align}
    \E{\EE{\Delta^2\est{\omega}(t)}}{p(N)} &\geq \E{\mrm{V}_{\infty}(t,N)}{p(N)} \nonumber \\
    &\geq \mrm{V}_{\infty}(t,\E{N}{p(N)}) = \mrm{V}_{\infty}(t,\bar{N})
\end{align}
where we have applied the Jensen's inequality since $\mrm{V}_{\infty}(t,N)$ is a convex function of $N$. Its convexity can be established by computing its second derivative w.r.t. $N$ and showing that it is positive $\forall N, t, q_\omega, \kloc, \kcoll$. Although computing the second derivative is lengthy, the positivity of all parameters ($N$, $t$, $q_\omega$, $\kloc$, $\kcoll$) assures that all the summands forming $\mrm{V}^{\prime \prime}(t,N)$ are also positive, and hence $\mrm{V}^{\prime \prime}(t,N) \geq 0$. To further support this analytical result, the convexity of $\mrm{V}_\infty(t,N)$ is also illustrated in \figref{fig:2Dplot_CSlimit}.

%%%%%%%%%%%%%%%%%%%%%%%%%%%%%%%%%%%%%%%%%%%%%%%%%%%%%%%%%%%%%%%%%%%%%%%%%%%%%%%%%%%%%%%%%%%%%%
%%%%%%%%%%%%%%%%%%%%%%%%%%%%%%%%%%%%%%%%%%%%%%%%%%%%%%%%%%%%%%%%%%%%%%%%%%%%%%%%%%%%%%%%%%%%%%
\end{appendices}
%%%%%%%%%%%%%%%%%%%%%%%%%%%%%%%%%%%%%%%%%%%%%%%%%%%%%%%%%%%%%%%%%%%%%%%%%%%%%%%%%%%%%%%%%%%%%%
\end{document}